\documentclass{ar-1col}

\usepackage[comma]{natbib}
\usepackage{url}
\usepackage{amssymb}
\def\aap{A\& A}
\def\aaps{A\&A Supp.}
\def\pasp{PASP}
\def\araa{AnnRevA\& A}
\def\mnras{MNRAS}
\def\nat{Nature}
\def\apj{ApJ}
\def\prl{PRL}
\def\aj{AJ}
\def\ao{App.Opt.}
\def\apjl{ApJL}
\def\apjs{ApJS}

\def\nar{New Astron. Rev.}

\jname{Xxxx. Xxx. Xxx. Xxx.}
\jvol{AA}
\jyear{YYYY}
\doi{10.1146/((please add article doi))}

\begin{document}

\markboth{Eisenhauer, Monnier \& Pfuhl }{Advances in Optical Interferometry}

\title{Advances in Optical / Infrared Interferometry}

\author{Frank Eisenhauer$^1$, John D. Monnier$^2$ and Oliver Pfuhl$^3$ 
\affil{$^1$Max Planck Institute for Extraterrestrial Physics, 85748 Garching, Germany; email: eisenhau@mpe.mpg.de}
\affil{$^2$Astronomy Department, University of Michigan, Ann Arbor, Michigan 48104, USA; email: monnier@umich.edu}
\affil{$^3$European Southern Observatory, 85748 Garching, Germany; email: opfuhl@eso.org}}

\begin{abstract}
After decades of experimental projects and fast-paced technical advances, optical / infrared (O/IR) interferometry has seen a revolution in the last years. The GRAVITY instrument at the VLTI with four 8 meter telescopes reaches thousand times fainter objects than possible with earlier interferometers, and the CHARA array routinely offers up to 330 meter baselines and aperture-synthesis with six 1 meter telescopes. The observed objects are fainter than 19 magnitude, the images have sub-milliarcsecond resolution, and the astrometry reaches micro-arcsecond precision. We give an overview of breakthrough results from the past 15 years in O/IR interferometry on the Galactic Center, exoplanets and their atmospheres, active galactic nuclei, young stellar objects, and stellar physics. Following a primer in interferometry, we summarize the technical and conceptual advances which led to the boosts in sensitivity, precision, and imaging of modern interferometers. Single-mode beam combiners now combine all available telescopes of the major interferometers for imaging, and specialized image reconstruction software advances over earlier developments for radio interferometry. With a combination of large telescopes, adaptive optics, fringe-tracking, and especially dual-beam interferometry, GRAVITY has boosted the sensitivity by many orders of magnitudes. Another order of magnitude improvement will come from upgrades with laser guide star adaptive optics. In combination with large separation fringe-tracking, O/IR interferometry will then provide complete sky coverage for observations in the Galactic plane, and substantial coverage for extragalactic targets. VLTI and CHARA will remain unique in the era of upcoming 30-40\-m extremely large telescopes (ELTs).

\end{abstract}


\maketitle

\tableofcontents

\newpage
\section{INTRODUCTION} 
\label{intro}

Optical and infrared (IR) interferometry is experiencing tremendous advances from leaps in sensitivity, precision, angular resolution, longer baselines, and better imaging.  

In this review, we present the technical and scientific achievements in optical/IR (O/IR) interferometry from roughly the past decade. We focus on the two main science-producing O/IR interferometers in the world, the European Southern Observatory's Very Large Telescope Interferometer (ESO VLTI), in particular GRAVITY, and the Georgia State University Center for High Angular Resolution Astronomy Array (GSU CHARA). In addition, the Large Binocular Telescope Interferometer (LBTI) and the Navy Precision Optical Interferometer (NPOI) are still in operation and will be discussed briefly.

Earlier reviews by \citet{quirrenbach2001} and \citet{monnier2003} thoroughly discussed the history of O/IR interferometry. More recently, a number of textbooks \citep{glindemann2011, labeyrie2014, buscher2015} have been introduced that augment the classic radio interferometry textbook from \citet{thompson2017}. We refer the reader to these sources for details beyond our cursory treatment.

In this introduction we will give a short history of the field, make comparisons to the more familiar radio interferometry, and motivate the reasons for the recent performance increases.

\begin{marginnote}[]
\entry{Astronomical interferometry}{Technique combining light from different telescopes to increase the angular resolution, taking advantage of the wave nature of light}
\entry{Optical / Infrared (O/IR)}{Wavelengths $\approx 0.4- \mbox{few 10 }\mu$m, with direct detection of photons, and observable from ground (other than far IR $\approx$ few ten - hundreds $\mu$m)}
\entry{Wavelength bands}{V: $0.5-0.6\,\mu$m \\R: $0.6-0.7\,\mu$m \\J: $1.1-1.4\,\mu$m \\H: $1.5-1.8\,\mu$m \\K: $2.0-2.4\,\mu$m \\L: $3.0-4.0\,\mu$m \\M: $4.6-5.0\,\mu$m \\N: $7.5-14.5\,\mu$m }
\end{marginnote}

\subsection{Short history of optical / infrared interferometry} 
\label{history}

The history of O/IR interferometry can be divided into three periods: The classical period from 1868-1930, the early-modern period from 1956-2005, and the modern ``present" day from $\sim$2000 to now. In this review, we propose that the 4th era has begun, one marked by GRAVITY and dual-beam phase-referencing on 8-10 meter (m) class telescopes that truly revolutionizes O/IR interferometry sensitivity.

\begin{figure}[t]
\vspace{-0.3cm}
\includegraphics[width=11cm]{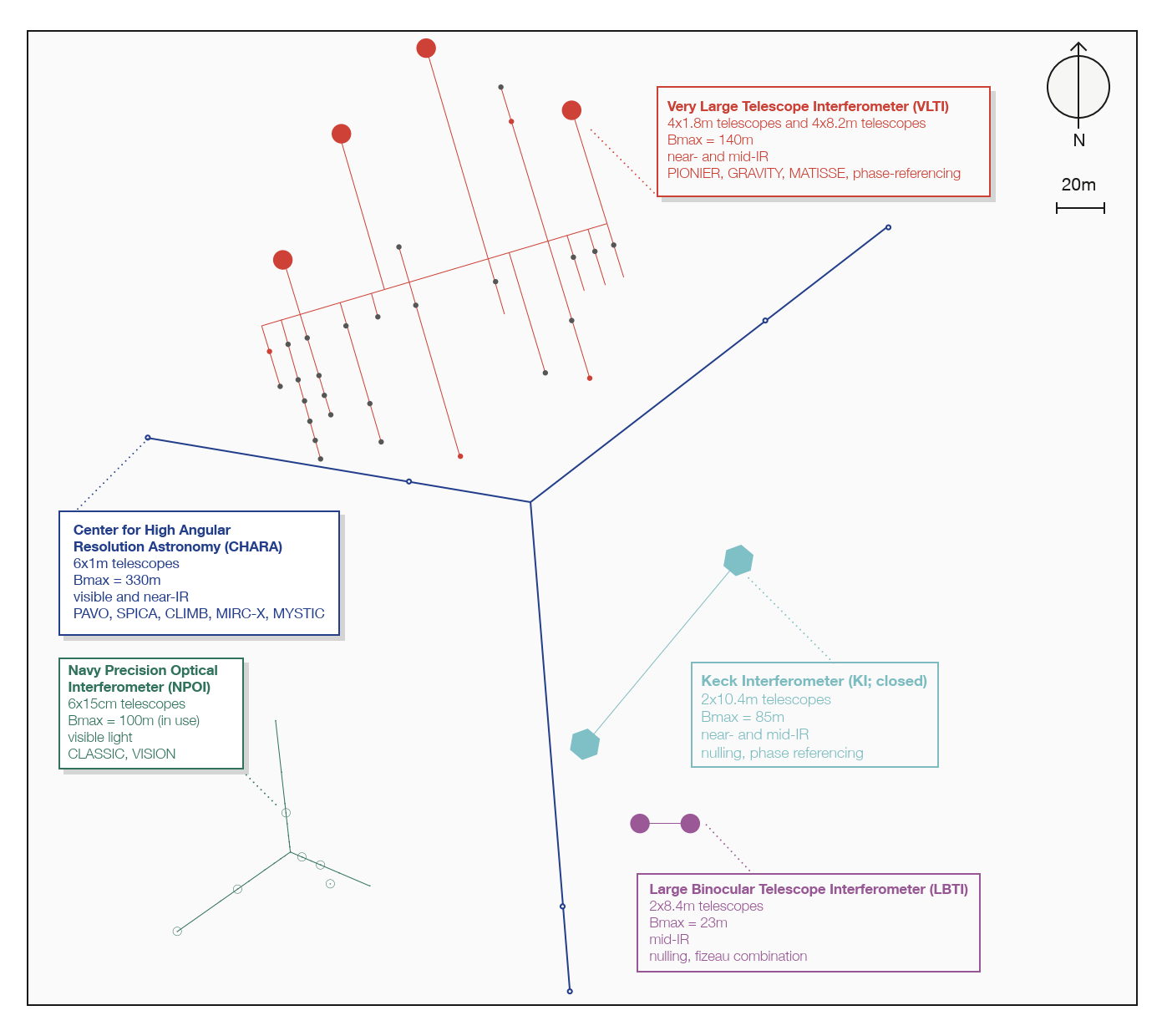}
\vspace{-0.3cm}
\caption{This sketch shows the footprints of modern O/IR interferometers, with telescopes and baselines to scale. The telescopes for LBTI (purple), CHARA (blue) and VLTI-UTs (red large circles) are fixed while the NPOI (green) and VLTI-ATs (red small circles) telescopes are mobile and can be re-positioned. }
\label{fig:facilities}
\end{figure}

The earliest period is rooted firmly in 19th century classical physics. \citet{fizeau1868} laid the foundation for how interference could be used to measure the sizes of stars by dividing the pupil of a telescope into small sub-apertures. The light from each pair of sub-apertures creates a ``fringe", the basic result of a Young's two-slit interference experiment. In this arrangement, each point source creates its own sinusoidal fringe pattern; thus, an extended object will create a low contrast interference as the peaks and troughs partially overlap to blur out the dark, destructive nulls present for point sources. Later, \citet{stephan1874} attempted on-sky experiments but was not able to resolve any stars. The effective angular resolution of an interferometer $\theta$ is defined as half the fringe spacing, $\theta=\frac{\lambda}{2B}$, expressed in radians for a wavelength $\lambda$ and a baseline between sub-apertures of $B$.    

\begin{marginnote}[]
\entry{Fringe}{Other word for interference pattern}
\entry{Angular resolution}{Smallest angular scale which can be well measured, typically full half width of a point source}
\entry{Diffraction limit}{Angular resolution $\frac{\lambda}{D}$ of a perfect telescope without aberrations and no blurring atmosphere}
\entry{Resolution of interferometer}{Given by the separation between telescopes, $\frac{\lambda}{2B}$ }
\end{marginnote}

Soon after Fizeau's and St\'ephan's pioneering work, Michelson developed a more thorough mathematical treatment, including coining the term ``fringe visibility" to describe the coherence -- with a visibility of unity corresponding to a perfect fringe with 100\,\% destructive interference at the fringe troughs and with a zero visibility for no interference at all. Michelson first deployed this method to measure the sizes of Jupiter's moons \citep{michelson1891}, eventually resolving Betelgeuse at 47 milli-arcseconds (mas) using a 20-foot interferometer on top of the Mt. Wilson 100" telescope \citep{michelson1921}. This ``classical" period ends with the largely-unsuccessful experiments to build the first truly long-baseline (50-foot) interferometer \citep{pease1930}.

The early-modern period of O/IR interferometry picks up with advances in electronics, optics, and detectors. The earliest revivals of long-baseline interferometry exploited new technologies, such as intensity interferometry \citep{hbt1956, narrabri1967} and heterodyne interferometry in the MIR by Nobel physicist Charles Townes \citep[e.g.,][]{johnson1974}. Following pioneering work by \citet{labeyrie1975}, early ``direct detection" interferometers, such as the Mark III \citep{markiii1988}, I2T \citep{i2t1985}, IRMA \citep{irma1993},  and others \citep{benedetto1983}, emerged and established the principles of our modern facilities, where light beams are collected at widely-separated telescopes, often transported through vacuum pipes, brought into coherence using moving delay lines, and finally interfered directly on a detector. 

Many of the projects in the 1980s and 1990s led directly to 2$^{nd}$ generation facilities: Narrabri intensity interferometer $\rightarrow$ SUSI \citep{susi1999}, McMath heterodyne interferometer $\rightarrow$ ISI \citep{hale2000}, I2T $\rightarrow$ GI2T \citep{gi2t1994}, Mark III $\rightarrow$ NPOI \citep{npoi1988} $+$ PTI \citep{colavita1999}, aperture masking $\rightarrow$ COAST \citep{coast1994}, IRMA $\rightarrow$ IOTA \citep{iota2003}. While all facilities from this era, except for NPOI, have been shut down, their collective technical impact has been impressive, setting the stage for the modern age.

\begin{marginnote}[]
\entry{Direct detection}{Detecting the photons e.g. by creation of photo-electrons in semi-conductor, also called homodyne detection}
\entry{Heterodyne detection}{Measuring the strength of the electric field by mixing with a local oscillator and measuring the power of the beating}
\end{marginnote}

The 3$^{rd}$,  so-called ``modern'' era starts with the debut of the ``flagship facilities" VLTI \citep{lena1979,beckers1990,vlti2007}, NASA Keck Interferometer \citep{colavita2013}, CHARA \citep{chara2005}, LBTI \citep{defrere2016} along with the evolution of NPOI. All of these facilities (except for the Keck Interferometer) are still operating as of 2022. See Figure\,\ref{fig:facilities} for the physical layout of these arrays. The only major facility under construction today is the Magdalena Ridge Optical Interferometer \citep[MROI;][]{mroi2013}, which aims at combining up to ten 1.4m telescopes over 350m baselines. Within the past decade, no new interferometers have come online. 

For a more historical perspective of O/IR interferometry, see the timeline explored by \citet{lawson2000} as well as the treatments by \citet{lena2020} and \citet{mcalister2020}. 

\subsection{Comparison with radio interferometry} 
\label{comparison}

\begin{marginnote}[]
\entry{Aperture masking}{Technique to recover diffraction limit of a telescope by placing a mask over the telescope, which only allows light through a small number of holes}
\entry{Speckle interferometry}{Technique to recover diffraction limit of a telescope by recording a series of short exposures}
\end{marginnote}

\begin{marginnote}[]
\entry{Radio}{Wavelength range from $\lesssim$ mm - meter,  observed by radio techniques, i.e. the amplification and mixing of the electric field  (heterodyne)}
\entry{Aperture synthesis}{Technique to produce images by interferometry with an array of telescopes}
\end{marginnote}

Even the largest, single radio telescopes of the 20$^{th}$ century barely reach the angular resolution of Galileo's first optical telescope from 1609. There is no practical way to build single telescopes large enough to sufficiently reduce the diffraction of the long radio-wavelengths. Therefore, ever since the development in the 1950s, aperture synthesis interferometry  \citep{jennison1958, ryle1960} has been the standard choice for high angular resolution telescopes in the microwave and radio bands \citep{thompson2017}. Why has this not been the case in O/IR astronomy? 

The principle of radio aperture synthesis interferometry is to measure the correlated flux from a celestial source on a number of baselines, each with two antennas, for different baseline length and orientation on sky. The source brightness distribution is then given by the Fourier transform. Each radio telescope is diffraction limited, with a flat wavefront across its aperture. With low-noise, phase-sensitive radio-amplifiers, all baselines can be measured simultaneously with little loss of signal or extra noise even for many telescopes. Since early radio interferometry operated at long wavelengths and narrow bandwidths, the coherence length is large, and the interference not much perturbed by the Earth atmosphere. As a result, the fringe can be maintained fairly easily over a long time and over a wide field of view. 

The situation in the O/IR is very different. Because of the short wavelengths, the product of coherence length and -time is $\approx 10^{6...8}$ smaller than in the radio. Other than in the radio, there are no low-noise heterodyne mixers and amplifiers in the O/IR, such that for multiple aperture interferometry the beams have to be split multiple times, resulting in large light losses. On the positive side, flux densities for thermal sources - like stars - are substantially larger for O/IR wavelengths, however, this advantage is eaten up by the need for multi-mirror, free beam propagation with much lower throughput. Taken together it is clear that O/IR interferometry is orders of magnitude more challenging than radio interferometry. 

This is especially the case when aiming for long exposures, which require to find at least one bright- and close-enough object to stabilize the fringes. While this is the case all over the sky for radio interferometers, we are very far from this situation in the optical because of atmospheric turbulence. But we are currently witnessing this transition for IR wavelengths with performance increase by large factors.

\begin{figure}[t]
\includegraphics[width=\columnwidth]{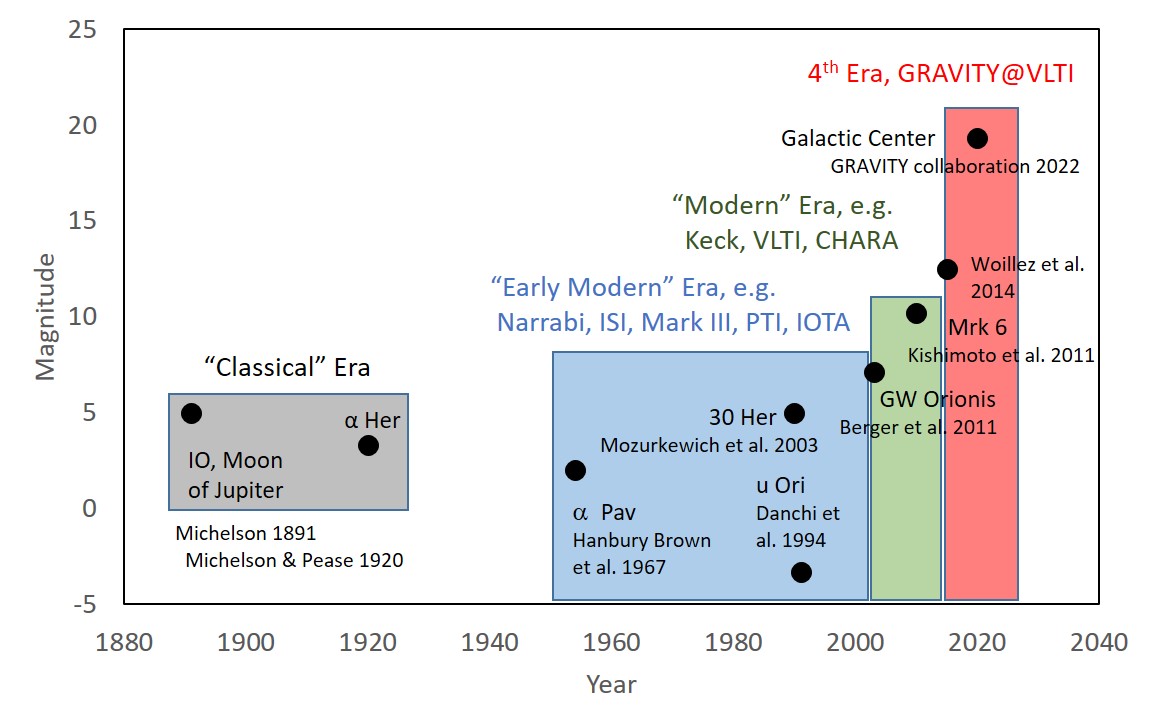}
\vspace{-0.5cm}
\caption{Sensitivity of O/IR interferometry from the ``classical" era with visual "by eye" measurements, followed by the ``Early Modern" and ``Modern" eras, to the latest advances from the ``4th" era with dual-beam phase-referencing. After a century of moderate improvement, the last decade has seen an increase by several order of magnitudes.}
\label{fig:sensitivity}
\end{figure}
\subsection{Performance increase by large factors} 
\label{perfromanceincrease}

We suggest a 4$^{th}$ era of O/IR interferometry has begun with GRAVITY and the advent and routine use of dual-beam phase-referencing on 8-m class telescopes (Figure \ref{fig:sensitivity}). The combination of adaptive optics (AO), dual-beam interferometry, and the ability to track the fast atmospheric fluctuations on a bright nearby star has allowed a breakthrough in sensitivity, extending coherent integrations from 50 milliseconds (ms) to $>$50 seconds (s), a 1000-fold jump. While the first demonstration was the early ASTRA experiment on the Keck Interferometer, the VLTI/GRAVITY project, initiated in 2005 by later Nobel Laureate Reinhard Genzel, has mastered the technique, and together with new detectors, integrated optics and improved laser metrology, allowed breakthroughs on the Galactic Center, exoplanets, and active galactic nuclei. 

\begin{summary}[INTRODUCTION - SUMMARY POINTS]
\begin{enumerate}
\item O/IR interferometry has matured and is offered at four major facilities.
\item VLTI and LBTI possess 8-m class telescopes for high sensitivity, CHARA and NPOI provide higher angular resolution and focus on imaging.
\item A new era of dual-beam phase-referenced interferometry has begun, allowing fringe detection on objects fainter than 19 magnitude with VLTI/GRAVITY. 
\end{enumerate}
\end{summary}

\section{ASTROPHYSICAL BREAKTHROUGHS} 
\label{breakthrough_astrophysics}

\subsection{Imaging of stellar surfaces} 

The current 4- and 6-telescope arrays have made interferometric imaging routine. Simple objects like binary stars have been imaged for some time but rarely offered advantages over model fitting. More challenging and scientifically-fruitful is imaging of stellar surfaces, where complex phenomena, such as convection and magnetic fields, play out and defy simply parameterization.

\begin{marginnote}[]
\entry{Interferometric Imaging}{see Aperture Synthesis, technique to make images with interferometers}
\entry{Red giant star}{Large and therefore luminous star in a late phase of its evolution, where the atmosphere is inflated and tenuous}
\end{marginnote}

\begin{figure}[t]
\includegraphics[width=\columnwidth]{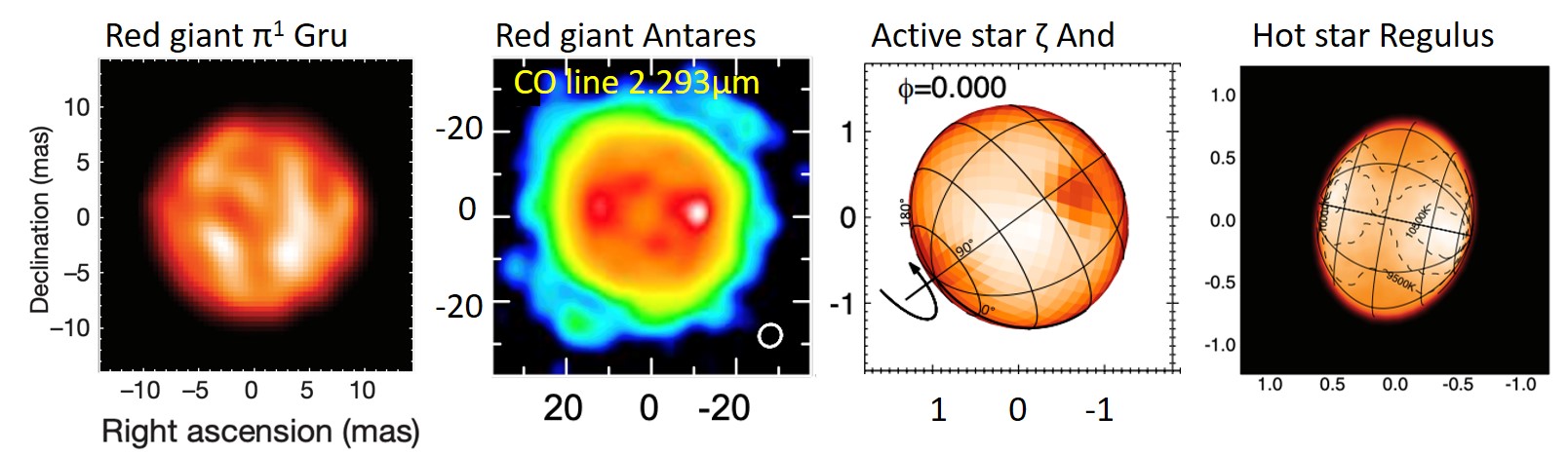}
\caption{The last decade has seen a breakthrough in stellar imaging as 4- and 6-telescope interferometers now have sufficient uv coverage and angular resolution to resolve the surfaces of red giants and supergiants, magnetically active stars, and even the closest hot rapid-rotators. Figures reproduced with permissions \citep[from left to right:][]{paladini2018,ohnaka2017,roettenbacher2016,che2011}.}
\label{fig:surfaces}
\end{figure}

Evolved stars host a number of non-trivial and interesting kinds of stellar physics -- pulsations, dust production, convective spots -- and these change on monthly to yearly time scales. Early imaging efforts \citep[e.g., IOTA;][]{haubois2009} suffered from sparse uv coverage due to the small number of telescopes and limited baselines available. More recently, efforts at the VLTI ATs and CHARA have led to remarkable images of red giants (Figure\,\ref{fig:surfaces}). They include state-of-the-art images by \citet{paladini2018} with approximately eight pixels across the photosphere, and rigorous interpretations inferred through 3D numerical simulations \citep[e.g.,][]{chiavassa2009}. With high spectral resolution, it is now possible to go beyond diameters and imaging of molecular shells \citep[e.g.,][]{perrin2004,perrin2020,lebouquin2009} and even {\em kinematically} resolve motions within molecular shells using, e.g, CO bandhead transitions \citep{ohnaka2017}. 

The long baselines of CHARA open up imaging for stars too far away or too small for other arrays. Sunspots caused by stifled convection from strong localized magnetic fields are seen on the Sun and inferred on other stars from photometric variations.  \citet{roettenbacher2016} and \citet{parks2021} published images of huge magnetic spots on all sides of the $\zeta$\,And and $\lambda$\,And systems respectively, finding asymmetric distributions of spots quite unlike Sun's. 

Lastly, sub-mas angular resolutions also allow imaging the bloated surfaces of the brightest, rapidly-rotating, hot stars \citep[e.g.,][]{vanbelle2012}, e.g., the surface of the B-star Regulus with a highly-oblate surface distorted by centrifugal forces and strong equatorial ``gravity'' darkening \citep{che2011}. Results on a half-dozen hot stars have led to new insights into non-spherical energy transport and advanced first-principle modelling of rotating stars \citep[e.g.,][]{lara2013}

\subsection{Revealing the inner astronomical units of circumstellar disks} 

One of the most fast-developing and exciting areas of astronomy today is planet formation. The mas angular resolution of O/IR interferometers translates into about 0.1\,Astronomical Units (AU) physical scale at nearby star forming regions, revealing the signposts of planet-formation, orbiting and outflowing dust, and complex physics of the star-disk connection which generated jets and outflows and transports angular momentum to the forming star \citep[see, e.g.,][for an overview]{dullemond2010}. IR interferometry  combined with ALMA's view of the outer disk \citep[e.g.,][]{dsharp2018} and O/IR coronagraphy of scattered light \citep[e.g,][]{benisty2022} is providing a rich and comprehensive picture of how planets are assembling in the earliest stages of their formation.

\begin{figure}[t]
\includegraphics[width=16cm]{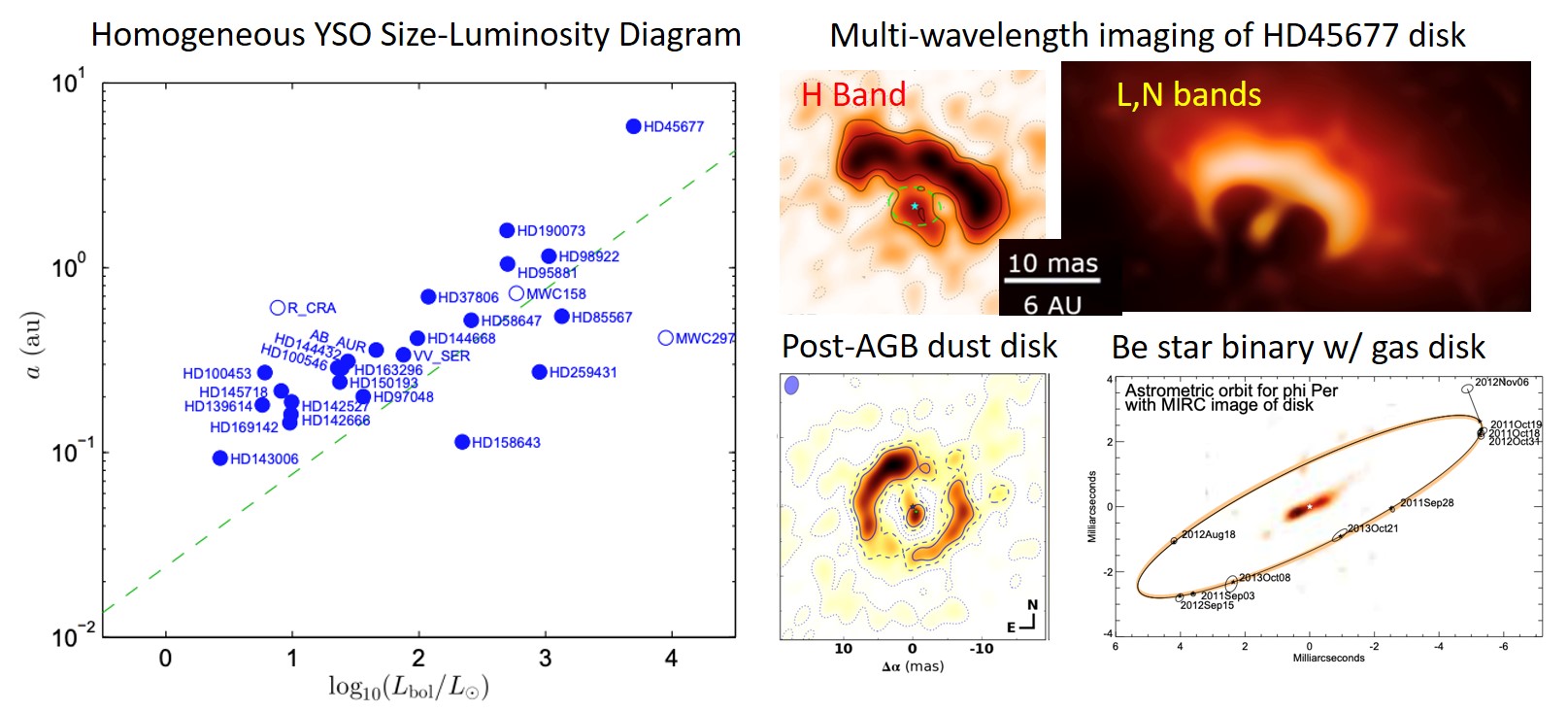}
\caption{IR interferometers now probe statistically-significant, homogeneous samples of YSOs, for example to study their size-luminosity relation (left).  Further, it is the only current technology that can image the inner AU of planet-forming disks, Be star gas disks, interacting binaries, and circumbinary disks. Figures reproduced with permission \citep{lazareff2017, kluska2020, lopez2022,  hillen2016, mourard2015}}
\label{fig:disks}
\end{figure}

The advent of sensitive 4-telescope combiners at VLTI have revolutionized studies of young stellar objects (YSO) by vastly increasing the number of objects observable, expanding wavelength coverage, improving homogeneity of the samples, and probing asymmetries using closure phases. 
\citet{lazareff2017} and \citet{perraut2019} measured homogeneous sizes and orientations of over dozens Herbig Ae/Be stars in H- and K-bands, vastly improving the data quality over earlier pioneering studies \citep[e..g,][]{monnier2002}. The size-luminosity diagram (Figure\,\ref{fig:disks}) shows a robust correlation over many orders-of-magnitude in luminosity, though with large scatter, potentially due to stochasticity from the formation of young planets. \citet{kluska2020} used advanced image reconstruction algorithms for some of these targets (HD\,45677 in Figure\,\ref{fig:disks}), finding only a few ring-like structures, with more often centrally-bright emission. Recent multi-epoch studies have further found a strong time variability in some objects \citep{kobus2020,sanchezbermudez2021}, the cause of which is unclear. A large sample of T\,Tauri disks in the K-band \citep{perraut2021} also found a strong breakdown of the size-luminosity relation \citep[in line with earlier Keck Interferometer results;][]{akeson2005}, related to greater importance of scattering and accretion. 

\begin{marginnote}[]
\entry{Young Stellar Object (YSO)}{Star in early phase of evolution, still surrounded by a circum-stellar disc} 
\entry{T\,Tauri star}{Low mass YSO ($< 0.2\rm\,M_\odot$)} 
\entry{Herbig Ae/Be star}{Intermediate -- high mass YSO ($2-8\rm\,M_\odot$)} 
\entry{H$\alpha$ and Br$\gamma$}{Hydrogen emission lines at 656\,nm and $2.16\,\mu$m}
\entry{CO band}{Band of molecular absorption or emission lines from Carbon monoxide, typically observed at K-band}
\entry{FU Orionis star}{YSO displaying extreme change of brightness and temperature} 
\entry{AGB star}{Asymptotic giant branch star - evolved, cool star, often creating circumstellar envelopes} 
\end{marginnote}

The Keck Interferometer and VLTI have also allowed to spatially and spectrally resolve the Hydrogen Br$\gamma$ line, probing the kinematics of accretion and the ``star-disk'' connection on sub-AU scales. Early studies did not find a clear picture, some disks showed compact Br$\gamma$ emissions smaller than the dust ring, while other showed emission on the same scales \citep[][]{kraus2008,eisner2009}. While only a few results have been obtained so far \citep[e.g.,][]{bouvier2020}, GRAVITY has the potential to carry out a large survey of Br$\gamma$ line emission as well as the CO bandhead \citep[e.g.,][]{garatti2020,wojtczak2022}. CHARA/VEGA \citep{perraut2016} was first to resolve the H$\alpha$ line in the accretion disk of AB\,Aur, finding a larger-than-expected extent from the magnetocentrifugal wind launched between the star and dusty disk's inner edge.

The shape of the dust sublimation front and contribution of gas/dust emissions very close to the star are too small to be definitively resolved by VLTI. CHARA, with $> 300$\,m baselines, has fully resolved the inner disk emission for two bright Herbig Ae/Be stars and found strong emission coming from inside the putative dust evaporation front \citep{setterholm2018}, corroborating earlier studies \citep[e.g.,][]{benisty2010}.
Recently, VLTI and CHARA data has been combined to directly image the rim shape for the Herbig Be star v1295\,Aql (Ibrahim et al, in press), finding a bright thin ring but with mysterious inner emission (see rightmost panel of Fig.\ref{fig:uvcov}).

New IR interferometry has validated some untested theory and solved long-standing mysteries.  \citet{labdon2021} found the disk temperature profile of the prototype FU\,Ori object to closely match  T$\propto r^{-\frac{3}{4}}$,  a 30-year old prediction by \citet{hartmann1985}. The Br$\gamma$ emission around TW\,Hya was interpreted as definitive proof of the magnetospheric accretion paradigm \citep{garcialopez2020}.  The complex dust geometry for the young interacting system GW\,Ori was finally solved by  \citet{kraus2020}. Lastly, \cite{labdon2019} and \citet{bohn2022} combined IR interferometry with adaptive optics to prove that inner disk misalignments produce the dark shadow bands seen at 100\,AU scales for many disks. 

Disks and outflows have also been imaged around other kinds of stars. 
\citet{mourard2015} were able to image the $\phi$\,Per disk with CHARA in both Br$\alpha$ and continuum (Figure\,\ref{fig:disks}), showing a clear connection between the disk geometry and close-in binary companion.  Circumstellar disks can also form in close interacting binaries \citep[Figure\,\ref{fig:disks},][]{zhao2008} and in post-AGB systems \citep[Figure\,\ref{fig:disks},][]{hillen2016}. The recent commissioning of the MIR combiner VLTI-MATISSE \citep[see early result by][in Fig.\ref{fig:disks}]{lopez2022} will revolutionize studies of dusty outflows in a wide variety of environments including AGB stars \citep[e.g.,][]{chiavassa2022}.

\subsection{Testing the black hole paradigm in the Galactic Center} 
\label{galacticcenter}

\begin{marginnote}[]
\entry{Black hole}{Object so massive and compact that not even light can escape, consequence of the General Relativity theory by Albert Einstein}
\entry{SgrA*}{Name of the radio source in the Galactic Center - the black hole}
\end{marginnote}

Motivated by the discovery of the first quasars in the 60s, \cite{lynden-bell71} proposed that most galactic nuclei, including the Galactic Center might host a supermassive black hole (SMBH). The discovery of the compact radio source SgrA* \citep{balick1974} at the core of the central nuclear star cluster provided some evidence for their proposition. However, SgrA* is faint in all bands other than the radio and sub-millimeter. With abundant gas in the inner 1\,parsec (pc) to fuel a potential SMBH, the case for an extremely underluminous SMBH was considered fairly unconvincing. This only changed with the advent of NIR Speckle and AO images of the central 1\,pc. Proper motions and later full orbits of stars demonstrated the existence of a compact central mass. The combination of precision astrometry better than 1\,mas with spectroscopy allowed to weigh the enclosed mass, to measure its distance and to set tight constraints on the density and therefore on the nature of the enclosed mass. By the end of the 2000s, the analysis of several dozen orbits in combination with radio measurements of the size and motion of SgrA* established that the radio source must be a massive black hole with about $4\times 10^6 M_{\odot}$, "beyond any reasonable doubt" \citep[for a review, see][]{genzel2010}. 

\begin{textbox}[]\section{2020 Nobel Prize in Physics for the Discovery of the Galactic Center Black Hole}
The Nobel Prize in Physics 2020 was awarded to Reinhard Genzel and Andrea Ghez for the "discovery of a supermassive compact object at the centre of our galaxy" and to Roger Penrose "for the discovery that black hole formation is a robust prediction of the general theory of relativity". The discovery of the Galactic Center is building on the experimental breakthrough in high angular resolution astronomy over the last 30 years, starting from Speckle Interferometry to recover the diffraction limited resolution of large telescopes in the 1990s, followed by AO and imaging spectroscopy in the 2000s, and initiated by Reinhard Genzel in 2005, with GRAVITY long baseline interferometry (since 2017) providing mas resolution imaging and few ten micro-arcsecond astrometry - the topic of this article. \end{textbox}

\begin{figure}[t]
\includegraphics[height=15cm]{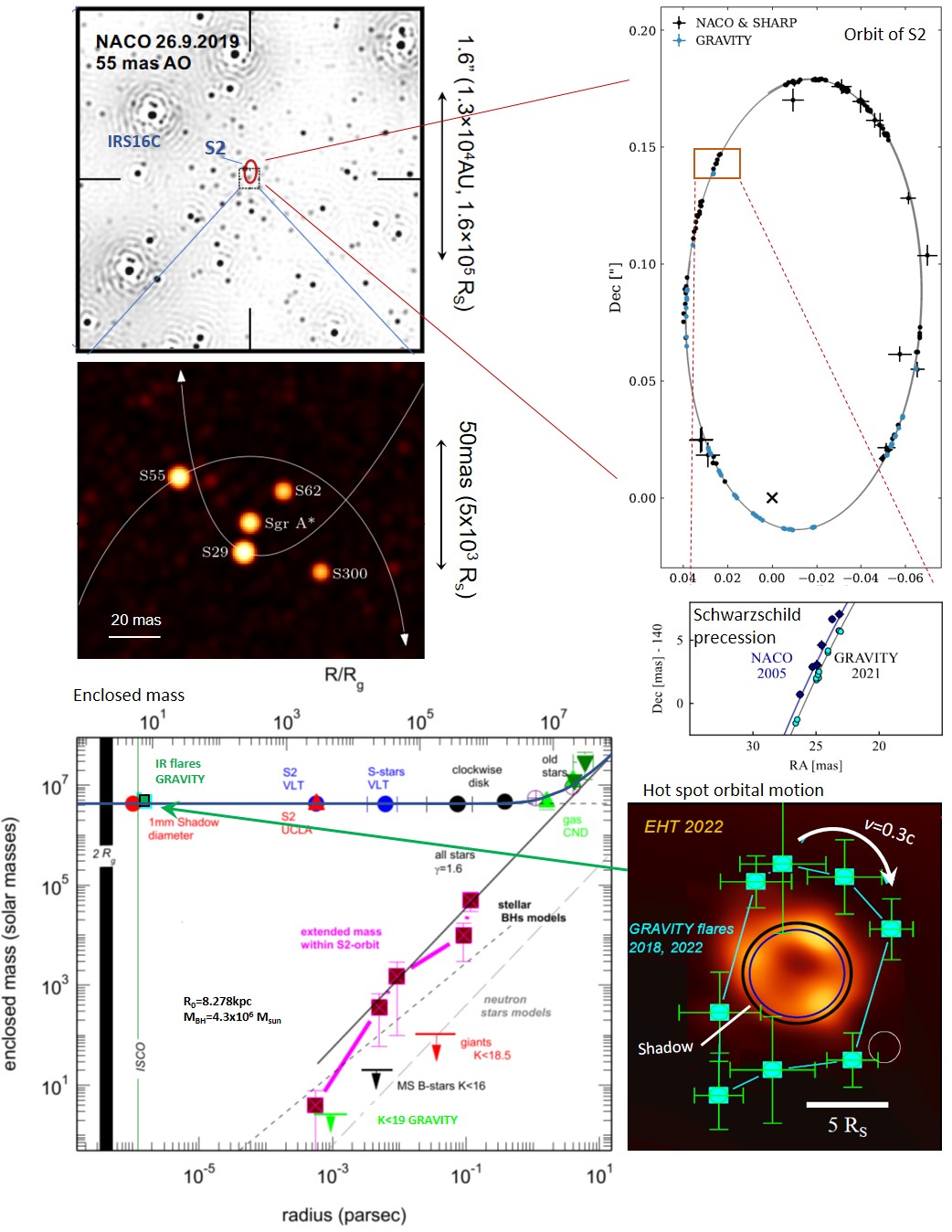}
\caption{Upper left: AO image of the Galactic Center obtained with an 8-m telescope. Middle left: Interferometric image of the central $\rm 0.1"$ \citep{gravity_gr,gravity_faint_stars}. The crowded region cannot be resolved with single telescope observations. Upper right: S2-SgrA* orbit.  \citep{gravity_schwarzschild}. Middle right: Zoom on the orbit of S2. The orbit does not close due to the Schwarzschild precession \citep{gravity_schwarzschild,gravity_mass_distribution}. Lower left: Enclosed mass in of the central 1\,pc centred on SgrA*. Lower right: IR flare positions observed over approx. 30\,min. The motion shows a  ($i\sim30^o$) orbit of a hot spot at 3-4\,$r_s$ \citep[][in prep.]{gravity_isco}. The background shows the EHT image of SgrA* \citep[from][]{EHT2022I}.}
\label{fig:GC}
\end{figure}

\begin{marginnote}[]
\entry{Event horizon}{The boundary of no escape. Its size is given by the Schwarzschild radius $r_s=\frac{2 G M}{c^2}$, where $G$ is the gravitational constant, $M$ the mass of the black hole, and $c$ the speed of light }
\entry{Innermost last stable circular orbit (ISCO)}{Closest orbit for particles, before general relativistic effects drag matter irrevocably into the black hole}
\entry{S-stars}{Population of young, high-mass stars orbiting close to the black hole}
\entry{Flares}{Sporadic emission at IR and X-rays, raising by factor few ten to hundreds above quiescent emission}
\end{marginnote}

The dynamical measurements from AO images allowed to derive the mass of the central object with a few percent accuracy. The motion of the stars follow almost perfect Keplerian orbits, even for stars like S2 that passes the central source as close as $1400\,r_s \approx \rm 120\,AU$ (Schwarzschild radius: $r_s=2GM/c^2$). Following the pericenter passage of S2 in 2002 it became clear that the first-order General Relativity (GR) effects will come in reach with precision observations \citep{rubilar2001}. Despite the fact that a black hole is a genuine prediction of GR, the signatures of GR on the stellar orbits with the leading post-Newtonian terms $(O(\beta^2), \beta=v/c)$, namely gravitational redshift and Schwarzschild precession, are small perturbations w.r.t. the Keplerian motion \citep[for a review, see][]{alexander2005}. The gravitational redshift scales with $\beta^2 \approx r_s/r$, in case of S2 approx. $\rm 200\,km/s$ compared to the maximum velocity of $\rm\,7700\,km/s$ at pericenter passage. The Schwarzschild precession rotates the elliptical orbit by $\approx$ 12.1' per 16\,yr revolution. In order to measure the effects, a significant (factor $4-10$) improvement in astrometry compared to what was possible in 2010 was needed. This posed one of the main science drivers for the development of the GRAVITY instrument \citep{eisenhauer2008,paumard2008}. Since the first light \citep{gravity_first_light2017}, GRAVITY has been regularly monitoring observing the central S-stars and SgrA*. On May 19, 2018, S2 passed pericenter with 2.6\,\% of the speed of light. By simultaneously monitoring the stars radial velocity and motion on the sky, \cite{gravity_redshift} were able to detect the gravitational redshift and transverse Doppler effect at high significance (later confirmed by \citealt{do2019}). The statistical robustness of the redshift detection was further improved by \cite{gravity_distance} to more than 20\,$\sigma$ significance. \cite{gravity_equivalence} used two atomic transition lines in the spectrum of S2 to test one pillar of the Einstein equivalence principle and thus General Relativity, the local position invariance (LPI). By separately measuring the redshift of the hydrogen and helium lines in the stellar spectrum, effectively two independent clocks can be probed, while moving through the black hole's gravitational potential. The results set an upper limit on a violation of the LPI of $5\times 10^{-2}$ for a change of potential which is six magnitudes larger than accessible with terrestrial experiments.

Only a few months after the pericenter passage of S2, GRAVITY captured several bright flares showing circular motion of the emission region \citep[Figure\,\ref{fig:GC}, ][]{gravity_isco}. The observed motion shows an orbit of a compact polarized “hot spot” of IR synchrotron emission at approximately 3 to 5 Schwarzschild radii of a black hole of 4.3 million solar masses. This corresponds to the region just outside the innermost, stable, prograde circular orbit (ISCO) of a Schwarzschild–Kerr black hole. The simultaneous motion, light curve and polarisation measurements of the flares allowed to constrain the inclination of the flaring region to a near face-on ($i\approx30^o$) orbit. The results are in remarkable agreement with the inclination and size later derived for the radio image of SgrA* \citep{EHT2022I}, suggesting that IR and radio emission originate both from the same region. The flare detection and the EHT image provide unique evidence that 4.3 million solar masses are contained in a region of a few $r_s$ (Figure\,\ref{fig:GC}, lower left), a mass density only explained by a black hole. \cite{gravity_flux_distribution} substantiated that SgrA* has two states: the bulk of the IR emission is generated in a lognormal process with a median flux density of 1.1\,milli-Jansky. This quiescent emission is supplemented by sporadic bright flares that create the observed power law extension of the flux distribution, and which are also observed in X-rays.

\begin{marginnote}[]
\entry{Gravitational redshift}{Spectroscopic signature that time slows down close to a black hole}
\entry{Schwarzschild precession}{Precession of elliptical orbits resulting from curved space time}
\entry{Einstein equivalence principle}{Outcome of any local non-gravitational experiment in free fall is independent of the velocity and its location}
\end{marginnote}

Two years after the detection of the gravitational redshift, \cite{gravity_schwarzschild} reported the detection of the prograde Schwarzschild precession induced by the gravitational field of the SMBH (Figure\,\ref{fig:GC}). The authors measured the mass of the black hole with 0.4\% accuracy and ruled out the presence of a binary SMBH. \cite{gravity_mass_distribution} refined the measurement and set an upper limit on an extended mass, e.g. a putative cusp of stellar remnants surrounding the SBMH, of less than $\rm 3000\,M_{\odot}$ within the apocenter of S2. 

The monitoring of the S-stars not only allowed to test the black hole paradigm but also allowed tackling a classical astrophysical problem; the distance of the sun from the Galactic Center. GRAVITY determined the distance between the Sun and the SMBH to $R_0 = 8277$\,pc with 0.4\% accuracy \citep{gravity_distance,gravity_schwarzschild,gravity_aberration}, confirming that the SMBH is located at the center of the Milky Way Bulge ($R_{0,bulge}=8210 \rm \pm80\,pc$, \citealt{bland-hawthorn2016}).

GRAVITY has delivered precision tests of Einstein’s general theory of relativity and the so far strongest experimental evidence that the compact mass in the Galactic Center (SgrA*) is indeed a Schwarzschild-Kerr black hole. What can we expect in the future? The upgrade of GRAVITY with its current sensitivity limit of $\rm K\approx19.5$ \citep{gravity_faint_stars} to $\rm GRAVITY+$ \citep{eisenhauer2019,gravity_wide} will push the sensitivity limit to $\rm K>22$, with the expectation to reveal more stars on even smaller orbits than S2. The astrometry from interferometry and the radial velocities from upcoming 30-40\,m telescopes will then allow to probe higher-order GR effects such as frame dragging of space time due to the spin of the black hole or the imprint of the black hole's quadrupole moment, and thereby might even provide a test of the general relativistic no-hair theorem. 

\subsection{Resolving the Broad Line Region and imaging the hot dust in Active Galactic Nuclei} 
\label{agn}

\begin{marginnote}[]
\entry{Active Galactic Nucleus (AGN)}{Center of a galaxy, for which the emission is dominated by the accretion on a massive black hole}
\end{marginnote}

An Active Galactic Nucleus (AGN) is a massive accreting black hole in the center of a galaxy with an Eddington ratio $L_{\rm AGN}/L_{\rm Edd} > 10^{-5}$, where $L_{\rm AGN}$ is the bolometric luminosity and $L_{\rm Edd}$ is the Eddington luminosity \citep[e.g.,][]{netzer2015}. AGN are thought to play an important role in galaxy evolution: energy released by AGN through radiation or powering outflows (i.e. AGN feedback) can transform star-forming galaxies into quiescent galaxies. The unified model of AGN assumes that a dust torus obscures the central engine, accretion disc, and the Broad Line Region (BLR), such that the AGN can only be observed directly from polar directions \citep{antonucci1985}.

\begin{figure}[t]
\includegraphics[width=11 cm]{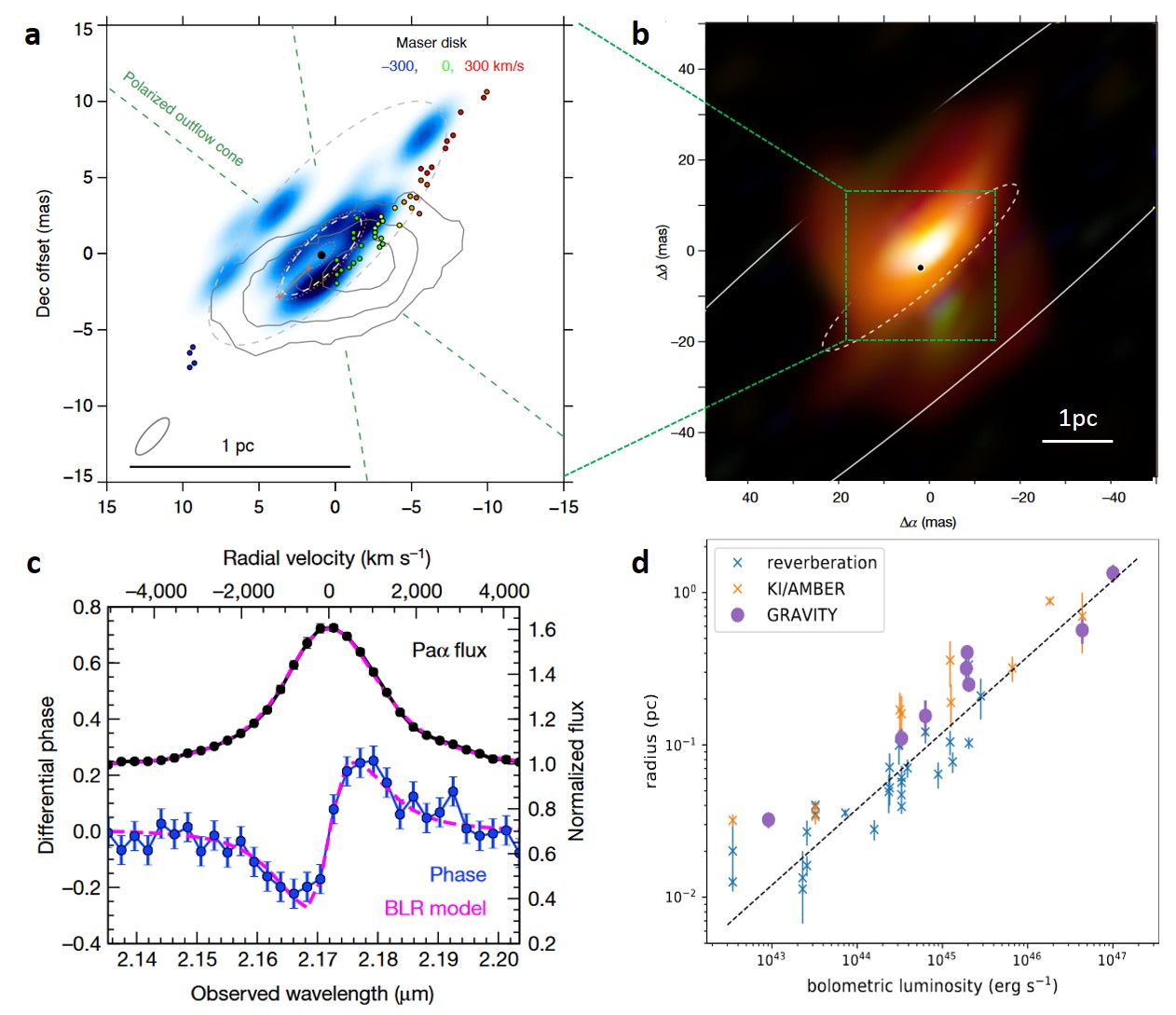}
\caption{(a): $K$-band image of the inner 2\,pc of NGC\,1068 \citep{gravity_ngc1068}. The dashed white ellipse corresponds to the dust sublimation radius. The filled black circle indicates the the AGN, and the kinematic centre of the masers (coloured circles); (b): MATISSE multi-color image of NGC\,1068 (credit: ESO/Jaffe, \citealt{rosas2022}); (c): \cite{gravity_3c273} spatially resolved the broad emission line kinematics of 3C 273. $\rm Pa \alpha$ line profile (black) showing non-zero phases and a change of sign across the broad emission line; (d): Radius-luminosity relation for dust size measurements (blue and orange crosses: from literature, filled circles: \cite{gravity_hot_dust}). The dashed line is the $R\sim L^{1/2}$ fit to reverberation measurements \citep{koshida2014}.}
\label{fig:AGN}
\end{figure}

IR interferometry has played a crucial role in the study of the torus region because the apparent size of 1\,pc at the distance of the closest AGNs ($\rm\sim20\,$Mega-parsec (Mpc)) is $\rm< 10\,mas$, a scale which can only be resolved with long baseline interferometry. While AGN are intrinsically bright in the IR, their relatively large distances require 8-m telescopes for observations. Early papers, using single baseline interferometers and $V^2$-fitting, identified the presence of multiple dust components with an elongated ($\rm 1.4\times0.5\,pc$) $\rm 600-800\,K$ dust core \citep[e.g.][]{jaffe2004}. About two dozen AGN have been partially resolved with the Keck interferometer \citep[e.g.][]{swain2003,kishimoto2011}, early VLTI \citep[e.g.][]{burtscher2013}, and more recently GRAVITY \citep{gravity_hot_dust,leftley2021}. This led to a dust size-luminosity relation for nearby AGN, independent of the relation inferred from dust reverberation mapping (Figure\,\ref{fig:AGN}d).

The advent of the 2$^{nd}$ generation VLTI instruments and the combination of four 8-m telescopes allowed for the first time to reconstruct images with GRAVITY and MATISSE. \cite{gravity_ngc1068} resolved the central 2\,pc of NGC 1068 in K-band with a spatial resolution of 3\,mas (Figure\,\ref{fig:AGN}a) and found a ring-like structure on sub-pc scales. The size matches that expected for the dust sublimation region, and the apparent orientation is similar to that of the maser disc, arguing for a common origin. This scenario is at odds with a geometrically and optically thick clumpy torus and instead argues for the presence of a dusty thin disc around the AGN, which is screened by dense and turbulent gas distributed on scales of 1–10\,pc, e.g. from AGN-driven outflows. This interpretation has been contested by \cite{rosas2022}, who resolved the central region with MATISSE at lower resolution but longer wavelengths from $3.7-12\rm\,\mu m$ (Figure\,\ref{fig:AGN}b). The derived dust temperatures and absorption values are consistent with a thick, nearly edge-on disk as predicted by the torus model. The different interpretation is largely driven by the assumptions made to align the radio continuum and maser emission and IR image.

\begin{marginnote}[]
\entry{Quasar (QSO)}{Extremely luminous AGN, with unobscured view on the central black hole and accretion disc, brightest objects in the universe}
\entry{Broad Line Region (BLR)}{Ionized gas clouds moving at high speed close to the black hole, observed as broad emission lines}
\entry{Dust Torus}{Opaque material in the equatorial plane obscuring the direct view on the central black hole}
\entry{Spectro-astrometry}{A differential measurement providing $\mu$as astrometry tracing the photo-center shift across an emission line}
\end{marginnote}

The BLR with an angular size $< 0.1\rm \,mas$ is even smaller than the hot dust region, and it is impossible to image even with the VLTI. Instead, the kinematics can be studied by ``spectro-astrometry", which measures the photo-center shift of the atomic gas as a function of wavelength (or velocity) across the emission line. The photo-center shift results in a small differential phase signal $<$\,1\,°, whose detection requires high sensitivity and deep integrations. \cite{gravity_3c273} for the first time detected the characteristic S-shaped phase signal of a rotating disk in the broad Pa$\alpha$ emission line of the quasar 3C\,273 (Figure\,\ref{fig:AGN}c). The signal is well described by a model \citep[following][]{pancoast2014} of fast moving gas clouds in a thick disk in Keplerian rotation around a supermassive black hole of $1.5-4.1\times10^8\rm \,M_{\odot}$. The inclination and position angles agree with those inferred for the radio jet. The measured emission radius is $R_{\rm BLR} = 0.12 \pm 0.03\rm \,pc$ (at an angular diameter distance of 548\,Mpc). To this day, three BLRs have been resolved successfully with spectro-astrometry \citep[3C 273, NGC 3783 and IRAS 09149-6206;][Figure\,\ref{fig:AGN}]{gravity_3c273,gravity_iras,gravity_ngc3783}, which revealed their structure, kinematics and angular BLR sizes with an unrivaled spatial resolution. The joint analysis of the angular BLR size measurement from GRAVITY and the linear BLR size from reverberation mapping campaigns allowed \cite{wang2020} to derive an angular distance of 3C273 of $552^{+97}_{-79}\rm \, Mpc$ and an independent measurement of the Hubble constant $H_0 = 72\rm\, km s^{-1} Mpc^{-1}$ with 15\% uncertainty. In a similar way, \cite{gravity_ngc3783_dist} found a geometric distance to NGC\,3783 of $39.9^{+15}_{-12}\rm\,Mpc$ and derived $H_0$ with a 30\% uncertainty.  GRAVITY already demonstrated first fringes of a redshift z=2.5 quasar \citep{gravity_wide}. Future BLR observations of a reasonably sized sample ($\approx30$ AGNs) will provide a new tool for measuring the masses of black holes at cosmological distances, and might allow to test the $H_0$ tension with $<3\%$ accuracy \citep{wang2020}. A similar tool might be provided by the interferometric dust parallax measurement as introduced by \cite{hoenig2014}. 

\subsection{Observations of exoplanets and spectroscopy of their atmosphere} 
\label{exoplanets}

While only applicable for a few dozen exoplanets so far, direct imaging offers the unique possibility to probe the thermal emission from the exoplanet's atmosphere, key for measuring the composition of their dense atmospheres. However, direct imaging is impaired by the small separation and the contrast between the exoplanets and their host stars, and only young, hot ($\lesssim 1000$\,K) and far ($\gtrsim 10$\,AU) planets are observable by AO and coronagraphy. IR interferometry has pushed both limits by orders of magnitude, providing the so far best-quality, high-resolution spectra from hot planets, orbit measurements with few ten micro-arcsecond ($\mu$as) astrometry, and first direct observations of planets previously known only from radial velocities. 

\begin{marginnote}[]
\entry{Exoplanet}{Planet outside the solar system}
\entry{Directly detected planet}{Emission from the exoplanet is directly seen in high contrast imaging or interferometry}
\entry{Radial velocity planet}{Exoplanet detected by the reflex motion and resulting radial velocity of its host star. Recognized with the 2019 Nobel Prize in Physics}
\entry{Transiting exoplanet}{Exoplanet which blocks some of the starlight as it passes in front of its host star}
\entry{C/O ratio}{Abundance ratio of Carbon over Oxygen, a tracer for planet formation history }
\entry{GAIA}{Space observatory measuring the positions, distances and motions of stars with unprecedented precision}
\end{marginnote}

\begin{figure}[t]
\vspace{0.5 cm}
\includegraphics[width=\columnwidth]{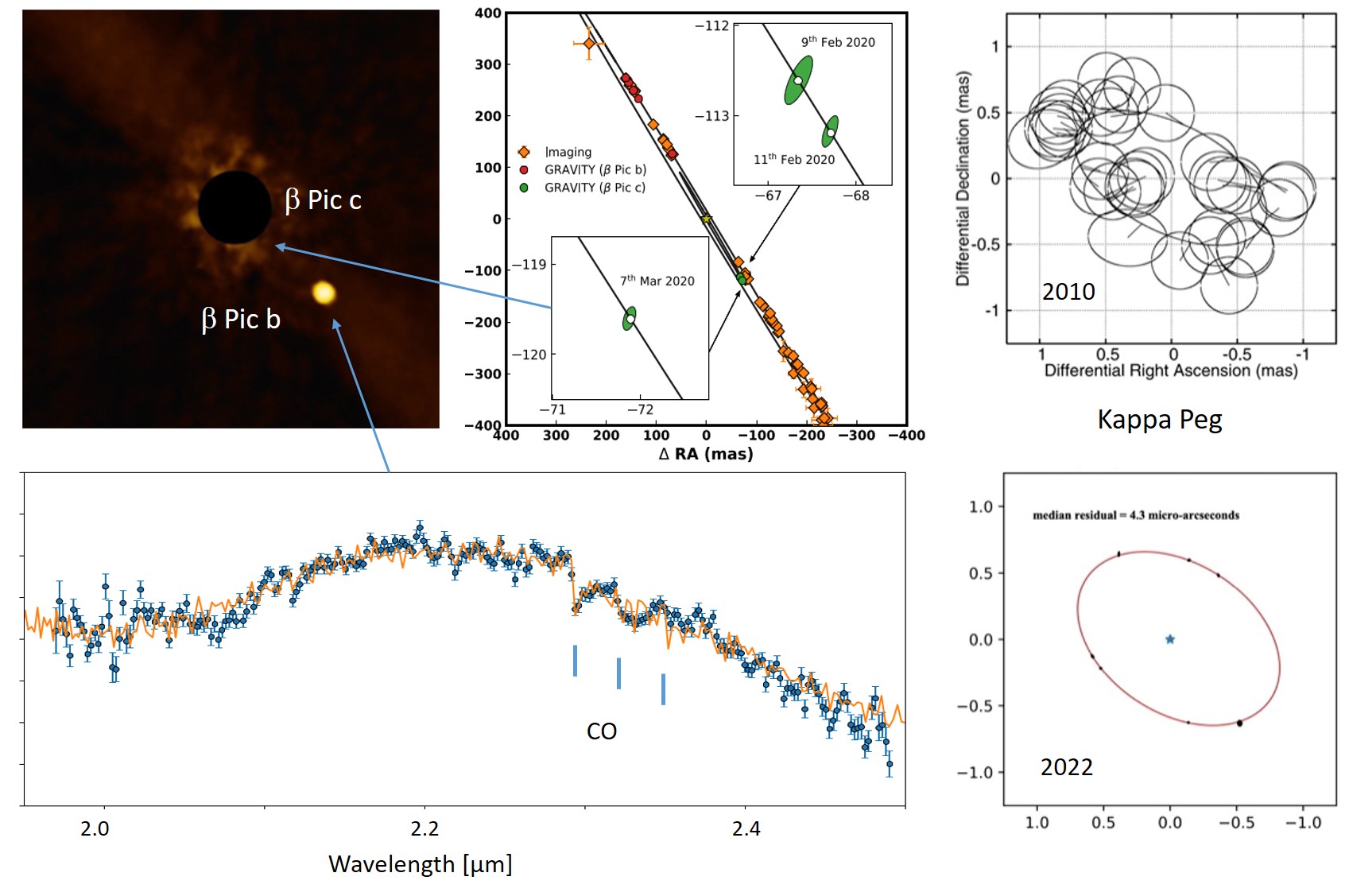}
\vspace{-0.5cm}
\caption{Interferometric observations of exoplanets: top left: AO image of the $\beta$\,Pic\,b exoplanet (credit: ESO/Lagrange/SPHERE consortium). The GRAVITY observations have resulted in the by far best spectrum of its atmosphere (bottom), and allowed to measure its C/O ratio, indicating that the planet has probably formed through core-accretion with strong planetessimal enrichment \citep{gravity_beta_pic_b}. The superior contrast and angular resolution of GRAVITY also led to the first direct detection an exoplanet, which was previously known only from radial velocity measurements \citep[$\beta$\,Pic\,c; ][]{lagrange2020,gravity_beta_pic_c}. The right panels illustrate the radical improvement in astrometry of binary stars over the past decade (top: astrometric wobble of $\kappa$\,Peg\,B from PTI \citep{muterspaugh2006} and CHARA \citep{gardner2021}, which should allow discovery of new exoplanets in the next years.}
\label{fig:exoplanets}
\end{figure}

The first detection of an exoplanet with interferometry \citep{gravity_hr8799} was HR\,8799\,e, a planet only 0.39\," from its host star. The spectra from GRAVITY are roughly ten-times higher signal-to-noise than possible with single telescope observations. This allows retrieving the properties of clouds and disequilibrium chemistry in the exoplanet atmosphere \citep{molliere2020} and to calibrate the mass-luminosity relations for protoplanets. Since this breakthrough, a series of observations have led to spectra for, e.g., $\beta$\,Pic\,b,c \citep{gravity_beta_pic_b,gravity_beta_pic_c,lagrange2020} and the PDS\,70 protoplanets \citep{wang2021}. The spectrum from $\beta$\,Pic\,b (Figure~\ref{fig:exoplanets}) allowed to peer into the formation history of this exoplanet: the low C/O ratio measured from the exoplanet spectrum, and the high mass of the exoplanet determined from astrometry, suggest a formation through core-accretion, with strong planetesimal enrichment.

The higher angular resolution and better contrast of interferometry has also led to first direct detections of exoplanets, which were previously known from radial velocity measurements, but which are too faint and too close to the host star for imaging with AO and coronagraphs. The first of these radial velocity planets detected and characterized with interferometry was $\beta$\,Pic\,c \citep{gravity_beta_pic_c,lagrange2020}, an 8--9\,$\rm M_{Jupiter}$ planet, orbiting at a distance of only 2.7\,AU inside the planet $\beta$\,Pic\,b discussed above. $\beta$\,Pic\,c is 11\,mag fainter -- a factor 25000 -- than the host star, and was detected at a separation as close as 96\,mas. The astrometric errors for the $\beta$\,Pic\,b,c planetary systems are as small as $20-50\,\mu$as. This allowed for the first time measuring the mass of an exoplanet from its gravitational imprint on the astrometry of another planet \citep{lacour2021}. 

One limitation for the direct detection of radial velocity planets by interferometry is the small field of view, and that the radial velocity technique does not provide the inclination of the orbit, therefore the direction to look for the planet. The $\beta$\,Pic planetary system is exceptional in this sense, because it is seen edge-on, both the debris disk as well as the orbit of the outer planet, thereby providing a good prior estimate for the location of the radial velocity planet. The second radial velocity planet directly detected by interferometry was HD\,206893\,c \citep{hinkley2022}, an $\approx$\,12\, $M_{\rm Jupiter}$ planet at the limit of the brown dwarf regime, maybe one of the rare planets exhibiting Deuterium burning in its center. In this case the detection of the radial velocity planet was guided by GAIA astrometry, which narrowed down the patrol field to be surveyed with the interferometer. Many additional planet detections are expected from GAIA astrometry \citep{wallace2021}, which will be reachable with interferometry, but not with traditional coronagraphic imaging. The GRAVITY+ upgrade \citep{eisenhauer2019} is expected to then see emission from $> 40$ gas giant exoplanets in the young co-moving groups close to the Earth, and an additional 30 exoplanets in more distant star forming regions.

Exoplanets have also long been sought using astrometry between two stars \citep{shao1992}. The GRAVITY collaboration has followed up this route with differential astrometry of young, nearby visual binary system, e.g. GJ\,65\,AB, WDS\,J20452-3120\,BC, and HD\,142\,AB, but results have not been published to date. The PHASES project on the Palomar Testbed Interferometer (PTI) measured differential phase between two close-by stars, achieving $\approx 100\,\mu$as precision \citep{muterspaugh2010}. This concept was updated for CHARA/MIRC-X and VLTI/GRAVITY observations using precision wavelength calibration and medium spectral dispersion to overlap fringe packets, and demonstrated $\approx 10-20\,\mu$as differential precision sufficient to detect giant exoplanets, though none have been reported so far \citep{gardner2022}.  

\subsection{Other major advances}
\label{otherresults}

\begin{marginnote}[]
\entry{Microlensing}{Gravitational magnification of a background star by another object passing in front, increasing its brightness. The lensed image is unresolved by single telescopes}
\end{marginnote}

There are many other notable firsts from O/IR interferometry in the past decade. \citet{dong2019,zang2020,cassan2022} were able to resolve the multiple images and arcs formed during a microlensing event. \citet{kraus2020b} used $\mu$as spectral-differential astrometry to measure the rotation axis of an individual star \citep{kraus2020b}. The high-mass x-ray binaries SS\,433 and BP\,Cru were probed with $\mu$as spectro-differential astrometry to resolve the gas and jet in these systems \citep{gravity_ss433,gravity_bp_cru}. \cite{kloppenborg2010} imaged the transit of a mysterious edge-on dusty disk across the face of the bright star $\epsilon$\,Aur. \citet{schaefer2014} watched Nova Del 2013 expand from 0.4\,mas on day 2 to $>10$\,mas a month later.

\begin{marginnote}[]
\entry{Microquasar}{Stellar mass black hole with mass accretion from a companion star, strong emission and jets, similar to supermassive black hole quasars}
\end{marginnote}

Interferometry is also used to measure fundamental properties of stars and our current facilities allow for extensive and rigorous surveys of stellar diameters as well as binaries. Here, we highlight the contributions  by  \citet{boyajian2012} to calibrate the effective temperature scale for main sequence solar-type stars, \citet{huber2012} to link precision diameters with asteroseismology using the sensitive visible-light CHARA/PAVO combiner \citep{pavo2008}, \citet{sana2014} to determine binary statistics for massive stars, \citet{gallenne2015} to measure masses for important distance-ladders Cepheids, \citet{montarges2021} to unveil the cause of Betelgeuse's recent dimming, and \citet{richardson2021} to measure the first dynamical mass of a N-rich Wolf-Rayet star using a binary separated by only $a=0.79$\,mas.

\begin{summary}[ASTROPHYSICAL BREAKTHROUGHS - SUMMARY POINTS]
\begin{enumerate}
\item Imaging the surfaces of stars with sub-milliarcsecond resolution -- including evolved stars, magnetic starspots, rapidly-rotating hot stars -- is now routine.
\item Warm and hot dust can be imaged around a large number of planet-forming disks and mass-losing stars, revealing unexpected dynamics and complexity.
\item Few 10 $\rm\mu arcsec$ astrometry of stars and gas as faint as $m_K=19$ allowed to test GR in the vicinity of the Galactic Center supermassive black hole.
\item Near- and mid-IR imaging of AGN allow testing the unified model. Spectro-astrometry of BLRs provides a new tool to measure black hole masses and distances. 
\item Superior contrast and angular resolution of interferometry produce better exoplanet spectra and orbits than AO coronagraphy.
\end{enumerate}
\end{summary}

\section{INTERFEROMETRY PRIMER} 
\subsection{Two telescope interferometer, angular resolution and field of view} 
\label{interferometry}

\begin{figure}[t]
\includegraphics[width=10 cm]{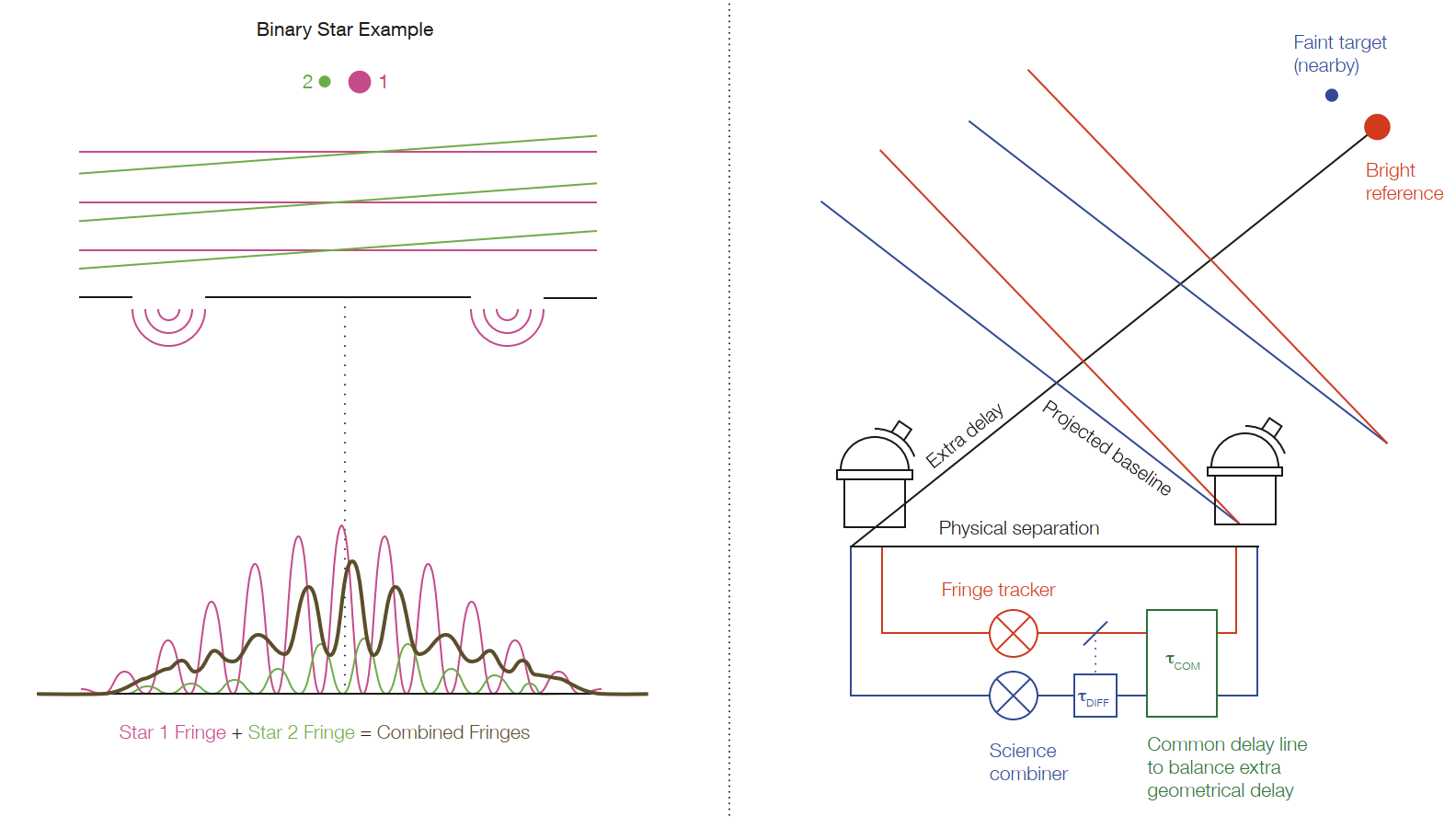}
\caption{Young's two-slit experiment (left panel) illustrates the basics of optical interferometry.  Each part of a distant source (red$+$green) will create its own fringe that will incoherently add together (black). The right panel shows the basic elements of a practical ground-based dual-beam interferometer.  Light from two fields are delayed together ($\tau_{\rm common}$) to account for common geometrical delay and then split to allow differential delay ($\tau_{\rm differential}$). The reference starlight (red) is detected at the fringe-tracking combiner and the derived atmospheric delays are then fed-back to the common delay line, allowing longer coherent integration on the science combiner (blue). This is analogous to a wavefront sensor/deformable mirror in a natural guide star AO system.}
\label{fig:young}
\end{figure}

Young's two-slit experiment illustrates the basic principles of interferometry (Figure\,\ref{fig:young}). When parallel wavefronts from a distant point source go through an aperture with diameter $D$, the light diffracts over a full-angle $\theta_{\rm beam}=\frac{\lambda}{D}$, where $\lambda$ is the observing wavelength. This angle is often referred to as the primary beam and typically sets the maximum field of view for most O/IR interferometers. When light goes through another aperture separated from the first by a baseline $B$, the electric fields interfere and produce a sinusoidal oscillation (``fringe") with a spacing of $\theta_{\rm fringe}=\frac{\lambda}{B}$ \citep{born1999}.

\begin{marginnote}
\entry{Primary Beam}{Diffraction-limited field of view of on telescope $\Theta=\lambda/D$}
\entry{Fringe Spacing}{Interference pattern projected onto sky $\Theta=\lambda/B$}
\entry{Spectral Resolution}{$R=\frac{\lambda}{\Delta\lambda}$}
\entry{Bandwidth-smearing field-of view}{ $\theta=R\frac{\lambda}{B}$}
\end{marginnote}

The number of fringes across the pattern is $\frac{B}{D}$, but this will be limited when using a broad spectral bandwidth. Then the number of fringes across an interferogram is set by the coherence length $\Lambda=\frac{\lambda^2}{\Delta\lambda}$ and will be equal to the spectral resolution $R = \frac{\lambda}{\Delta\lambda}$. If $R$ is too small, fringes will not completely fill the diffraction pattern and this will further restrict the effective field of view to $\theta\sim\frac{\lambda}{B} R $. The coherent field of view matches the telescope diffraction limit for a spectral resolution $R=\frac{B}{D}$, which is typically the lowest fractional bandwidth used in practical interferometers. 

\subsection{Complex visibilities} 
\label{vanCittertZernike}

\begin{marginnote}[]
\entry{Fringe visibility $\tilde{V}$}{Complex number representing amplitude and phase}
\entry{(u,v) point}{The projected baseline (East, North) of the interferometer as seen from the object in units radians$^{-1}$, associated with each visibility point}
\end{marginnote}

When an interferometer looks at a complex object, the interference pattern changes with respect to that of a point source. Figure\,\ref{fig:young}a illustrates the basic principle.  Each point of light on the distant source will create its own interference fringe slightly shifted with respect to each other, since the wavefronts are slightly tilted, and with an intensity proportional to the brightness of each part of the source. Since astronomical sources are incoherent, these two sine waves add up in power resulting in a sine wave that is characterized by a fringe amplitude and phase. The normalized amplitude of the fringe is called the contrast or the ``visibility" $V$ where unity means the fringe is fully-modulated, creating dark destructive nulls and bright constructive peaks; a zero visibility shows no modulation across the interferogram. O/IR interferometrists have historically used this normalized visibility since the total flux is often poorly measured. Radio astronomers use the coherent flux instead -- the total flux times the normalized visibility -- with units of flux density (W/m$^2$/Hz), and this practice is becoming common for some IR applications when the normalization is problematic, for example in the thermal IR. The relative position of the fringe gives the ``phase" $\phi$ of the interference. The combination of the two quantities visibility amplitude $V$ and phase $\phi$ forms the ''complex visibility" $\tilde{V}=e^{i\phi}$.

Interpreting this complex visibility is straightforward. Since each point on the sky creates a sine wave, the final observed complex visibility $\tilde{{V}}$ is simply an integral of sky intensity times sine waves, otherwise known as a Fourier Transform. This relationship is captured by the van Cittert-Zernike Theorem:
$\tilde{{V}}(u,v) \propto \int_{\rm sky} I(\alpha,\delta) e^{-2\pi i (u \alpha +v \delta) } d\alpha d\delta$, where $I(\alpha,\delta)$ represents the target brightness distribution on the sky as function of right ascension ($\alpha$) and declination ($\delta$) and $(u,v)$ represents the projected baseline in (east,north) components.

\subsection{Atmospheric coherence lengths and times}
\label{atmosphere}

\begin{marginnote}[]
\entry{Fried parameter $r_0$}{Diameter for which the root-mean- square (rms) wavefront error introduced by the atmosphere is 1\,rad}
\entry{Coherence time $\tau_0$}{Time span for which the rms phase fluctuations from the atmosphere are 1\,rad}
\entry{Coherence length}{Path difference  $\Lambda=\frac{\lambda^2}{\Delta\lambda}$} between the telescopes, for which the fringe contrast drops to 0.5
\entry{Isoplanatic angle $\theta_0$}{Angular separation, for which the phase difference of two objects fluctuates by 1 rad rms}
\end{marginnote}

In order to translate the elegant principles of interferometry to a practical facility, we must also account for properties of the atmosphere. In this section we will just introduce the basic picture with further elaboration in \S\ref{breakthrough_sensitivity}, and refer the reader to a more in-depth treatment by \citet{quirrenbach_mss2000}.

The ideal picture of an interferometer starts to break down when we consider light propagating through the atmosphere. The perfectly flat wavefronts at the top of the atmosphere become distorted as they encounter varying densities of air. Simplistically, we can follow each ray and add up the time delay caused by the index of refraction of air. Rays close together go through essentially the same air and so have small root-mean-square (RMS) variation while rays far apart become more different. The Fried parameter $r_0$, or atmospheric coherence length, is defined as the diameter of the circular aperture that has an average RMS wavefront error of 1 radian in phase. This value is wavelength dependent both because the phase depends on wavelength and the index of refraction varies some with wavelength. The theory of Kolmogorov turbulence predicts that $r_0 \propto \lambda^\frac{6}{5}$, and that telescopes with diameters $D > r_0$ will have long-exposure images with seeing limited angular resolution $\theta_{\rm seeing}=\frac{\lambda}{r_0} \propto \lambda^{-\frac{1}{5}}$. Thus, seeing-limited image quality improves into the IR counter to the diffraction limited performance.
Sites with excellent seeing will have $r_0\,\approx 20$\,cm in the visible (500\,nm) and $r_0\,\approx 1$\,m at K band (2.2$\mu$m). As explained later, the limitation of $r_0$ can be overcome by including AO.

Most observations in astronomy have long exposure times, minutes or even hours. For interferometry, this is only possible when actively stabilizing the fringes to well better than a wavelength, a technique called fringe-tracking (\S\ref{fringetracking}). Otherwise, the turbulent atmosphere causes time-varying optical path lengths above each telescope and a changing phase of the interference fringe. As the phase changes, the troughs and peaks will blur together in a long exposure ruining the measurement. Thus, interferometers must observe fringes with short exposure time to ``freeze" the atmospheric motions. It is useful, though incomplete, to adopt Taylor's frozen atmosphere hypothesis that assumes the wavefront errors are fixed as they are blown across the telescope aperture at a wind velocity $v$. If so, then the typical coherence time is given by $\tau_0 = \frac{r_0}{v}$. For upper atmosphere speeds of 10\,m/s, we find $\tau_0 \approx 20$\,ms in the visible at excellent sites, though jetstream speeds of $> 50$\,m/s can drastically reduce the coherence time to $< 3$\,ms. Again, note the coherence time will be much improved in the IR compared to visible due to the dependence on $r_0$.

Combining $r_0$ and $\tau_0$ leads to the concept of a ``coherent volume" of photons that can be used for estimating sensitivities. Without AO and fringe-tracking, the largest useful aperture is $r_0$ and the longest coherent integration time is $\tau_0$, thus the photon volume is proportional to $r_0^2 (c \tau_0) \propto \lambda^{3.6}$, according to Kolmogorov turbulence. This strong dependence on wavelength explains why IR interferometry has been much more developed than visible light interferometry. \S\ref{breakthrough_sensitivity} will show how advanced methods now can largely overcome this traditional limitation to interferometry sensitivity, boosting fringe sensitivity by over $\times$1000 in the past 10 years.

\subsection{Practical implementation}
\label{implementation}

\begin{marginnote}[]
\entry{Delay line}{Device to delay light from one interferometer arm compared to another, typically to account for geometrical delays}
\entry{Beam combiner}{Instrument to bring light from separate telescopes together in order to measure mutual coherence}
\entry{ABCD}{Shorthand for measuring interference at four phases 0, $\pi$/2, $\pi$, 3$\pi$/2 in order to determine mean power, fringe amplitude, and phase}
\end{marginnote}

While the monolithic binocular mount of the LBTI resembles Young's two-slit configuration, most interferometers are an abstraction of this experiment. The apertures are replaced by telescopes which must be corrected by AO or be limited to a size $\approx r_0$. Once corrected, light from each telescope is sent to the delay lines. The wavefronts for each telescope must be dynamically delayed based on the sky position of the target since wavefronts arrive obliquely, intercepting some telescopes before others. In order to measure an interference fringe, the optical path difference between telescopes must be stabilized either within a coherence length using group-delay tracking or within 1~radian by phase tracking, the latter allowing long coherent integrations. For wide-field phase-referencing (separation $> 10$\,"; only at VLTI), the light from the reference and science targets must be split at the telescope and sent through the delay lines on separate beams. Along the way, we must seek to minimize any differential dispersion and birefringence, while maintaining high transmission. With typically more than 20 surfaces from the telescope up to the instrument, the beamtrain transmission alone is typically low ($< 20$\,\% in IR, $< 10$\,\% in visible). Once the optical path differences and differential polarization (e.g., by half- and quarter-wave plates) have been compensated, all the beams can now be interfered. Figure\,\ref{fig:young} (right panel) illustrates these basic components of today's interferometers.  

Beam combiners fall into two general categories: ``all-in-one" or ``pairwise". As the name implies, an all-in-one combiner will overlap light from several telescopes together, creating many fringes simultaneously. To distinguish different baselines, the interference patterns are either coded spatially (e.g., in image-plane) or modulated temporally. In a pairwise combiner, the light is split up and combined in pairs, so each combiner only has two input beams. Also here, the interference can be scrutinized with either temporal-modulation using a beamsplitter or through a so-called ``ABCD" combiner, which further splits the beams and adds an extra combiner to allow four quadrature fringe phases to be measured simultaneously. The ABCD quadrature can also be achieved by temporal modulation and reading the detector at four fixed phase shifts. CHARA/MIRC-X is an example of an all-in-one combiner, VLTI/GRAVITY is an example of a pair-wise combiner using the ABCD method, and NPOI/CLASSIC combiner combines aspects of both.

In order to improve calibration of fringe coherence, most combiners apply some kind of ``spatial filtering'' to remove aberrations from the incoming beams before interference.  The purest way to do this is using single-mode waveguides such as small-core fibers or planar waveguides, although pinholes can also be used (e.g., VLTI/MATISSE) when no fibers are available. During this process, the single coherent flux can fluctuate wildly with atmospheric turbulence, although AO improves this. For calibration, the coupling fraction into the single-mode must be monitored in real time. The FLUOR combiner on IOTA \citep{foresto1998}
pioneered the use of single-mode waveguides for interferometry and many combiners today now are based on this breakthrough (NPOI/VISION, CHARA/FLUOR+SPICA+MIRCX+MYSTIC, VLTI/PIONIER+GRAVITY). In \S\ref{singlemode} we discuss how the miniaturization made possible by telecommunication technologies have enabled new generation of sophisticated instruments.

\subsection{Closure phases, phase-referencing, and fringe-tracking} 
\label{closure}

While a fringe measurement reveals an amplitude and a phase, the fringe phase is initially corrupted by atmospheric turbulence.  A time delay $\Delta \tau$ above one telescope due to change in air density along the path of the photons will shift the phase of the fringe by $\phi_{\rm rad}=2\pi\frac{c\Delta \tau}{\lambda}$ and this changes by 1\,radian every coherence time $\tau_0 \ll 1$\,s.  The phase excursions are too large and the statistics not sufficiently stationary to allow long term averaging. Two-telescope interferometers only average the $|\tilde{{V}}|^2$ and fit models to estimate stellar diameters or binary parameters. Complex astrophysical objects encode much information in the Fourier phases and so there is the need to calibrate these atmospheric fluctuations in order to carry-out true imaging with interferometers.

\begin{marginnote}[]
\entry{Closure Phase}{Phase observable immune to atmospheric turbulence, made from the sum of three fringe phases measured around a closed triangle of baselines}
\entry{Group delay tracking}{Maintaining the optical path difference to within one fringe packet coherence length $R \lambda$}
\entry{Phase tracking}{Maintaining the optical path difference to within 1~radian of phase delay $\frac{\lambda}{2\pi}$}
\entry{Dual-beam phase referencing}{Technique to allow long coherent fringe integrations of a faint target by real-time fringe tracking of a bright nearby reference star}
\entry{Baseline bootstrapping}{The technique of fringe tracking on two short baselines in order to coherently average on a longer baseline}
\end{marginnote}

Radio interferometry encountered this phase instability problem first and introduced the concept of {\em closure phase} \citep{jennison1958}, which is an observable quantity for a triangle of baselines and which is immune to phase instabilities. Any fringe phase shift related to a single telescope (not baseline) can be removed by adding up fringe phases in a closed triangle.  This can be seen by realizing that phase shift at telescope 2, for instance, will cause an equal but opposite-signed shift for baseline 1$\rightarrow$2 and baseline 2$\rightarrow$3. Closure phases thus reveal linear combinations of the true fringe phases we need for modeling and image reconstructions but not all the information.  Numerically, the number of independent phases in an $N$ telescope interferometer is $N (N-1)/2$ while the number of independent closure phases is $(N-1)(N-2)/2$. For example, a 3-telescope array accesses one closure phase out of three phases, a 6-telescope array (CHARA) access ten out of 15 phases, and 50-telescope array (ALMA) measures 1176 out of 1225 phases. For more detailed explanations and examples, see the thorough treatment  by \citet{monnier_mss2000}. Here we summarize a few important properties of closure phases compared to Fourier phases. Firstly, astrometric information is lost within closure phases since an image shift on the sky is equivalent to adding a planar geometrical delay above each telescope, thus cancels out in a triangle. Point-symmetric distributions (like uniform stars or simple inclined disks or rings ) have phases of 0 or $\pi$ when the origin is centered on the object, thus closure phases of point-symmetric objects also can only be 0 or $\pi$.  Closure phases provide absolutely crucial information for asymmetric objects and can be interpreted with forward modeling (\S\ref{imaging}).

While closure phases allow recovery of phase information, their use does not extend the coherence time and thus not overcome the sensitivity limitations from short exposures. ``Phase referencing" can be used to allow much greater sensitivity by using a bright nearby star as a phase-reference, a kind-of AO for interferometry that is used extensively in the radio. This technique has limits since one has to find a reference star within the ``isopistonic" patch, which is the patch of sky that sees the same turbulence; the wavefronts from stars more than $\approx 30$\," away (in NIR) encounter different turbulence and thus the phase errors are mostly uncorrelated. While phase-referencing was demonstrated on the Keck Interferometer shortly before being shut down \citep{woillez2014}, the VLTI/GRAVITY instrument was the first to make the technique practical and is leading us into a new era of high sensitivity observations -- see science highlights in \S\ref{breakthrough_astrophysics} along with more technical details in \S\ref{breakthrough_sensitivity}.

More common than dual-beam phase-referencing is fringe-tracking using self-referencing. Here, one uses part of the light of the target to track the fringes while reserving some of the light for ``science'' observations.  For instance, at the Keck Interferometer, light from one side of the beamsplitter was sent to the low-resolution spectrograph for high sensitivity fringe tracking, while the light from the other side of beamsplitter was sent to a high resolution spectrometer \citep{woillez2012}.  Often one wavelength channel is used for real-time fringe-tracking to allow integration at another wavelength \citep[e.g, Keck Nuller, VLTI/FINITO, CHARA/CHAMP;][]{serabyn2012,gai2003,champ2012}.  The VLTI/GRAVITY fringe tracker \citep{lacour2019} allows long integrations at K-band or with the MATISSE instrument in L-, M- and N-bands. Note that the fringe phases do not need to be tracked on all baselines since a long-baseline fringe (say between telescopes $1-3$) can be stabilized as long as fringes of two connecting shorter baselines ($1-2$ and $2-3$) are available -- a technique called ``baseline bootstrapping.''

\subsection{Homodyne, heterodyne, intensity, and quantum-enhanced interferometry} 
\label{alternatives}

We conclude this section with a short summary of different kinds of interferometry under active development.  In \S\ref{intro}, we already introduced the main technique of ``direct detection" or ``homodyne" detection where the light from the different telescopes is interfered before detection by a square-law detector, which measures the intensity = $|{\rm amplitude}|^2$.  This is by far the most common way optical interferometry is carried out today.  

\begin{marginnote}[]
\entry{Homodyne interferometry}{Electric fields from two beams are mixed together before being measured by a square-law detector}
\entry{Heterodyne interferometry}{Electric field is mixed with a laser local oscillator (LO) at each telescope, the output of the square-law detector maintains the original phase information. Signals can subsequently be interfered using standard radio techniques.}
\entry{Quantum noise}{Heterodyne detection introduce extra noise related to Poisson fluctuations from the local oscillator signal}
\entry{Quantum local oscilator (QLO)}{Uses an entangled photon source to eliminate shotnoise in heterodyne interferometry}
\end{marginnote}

A modified version of this method involves up-converting the light at each telescope to higher frequency using non-linear optics \citep[e.g.,][]{boyd1977,aloha2016}.  This upconverted light can then be interfered using traditional direct detection methods, potentially with advantages in the thermal IR where fiber optics and detectors are inferior to shorter wavelength devices.

Heterodyne interferometry \citep{johnson1974,hale2000} has been a practical method requiring only local measurements at the telescopes by mixing with a carrier signal and with interference happening separately. Often, the information can be recorded with correlation happening ``in software," or  using special-built digital correlators. Heterodyne interferometry in the visible, NIR, and mid-infrared (MIR) is intrinsically much less sensitive than direct detection as it involves mixing the starlight with a bright laser as a local oscillator, introducing laser shot noise as a background that becomes severe shorter than $20\,\mu$m. See \citet{ireland2018} for recent comparison of heterodyne and homodyne sensitivity in the context of the MIR Planet Formation Imager concept and \citet{berger2020} for plans to radically increase bandwidths using multiplexing.

As already mentioned, intensity interferometry measures correlations in photon arrival times at different telescopes and is one of the earliest true ``quantum optics'' discoveries \citep{hbt1956}. This correlation requires two photon processes and thus is intrinsically less sensitive than homodyne methods, though again can be possibly practical for some applications.  Attempts to revive this technique are underway taking advantage of the large collecting area and baselines from current and next-generation Cherenkov gamma ray telescopes \citep{nunez2012,cta2019} and advances in detectors for extreme wavelength multiplexing \citep{horch2013}

The latest frontier in optical interferometry to open up is using quantum resources, such as entangled photons, to confer a ``quantum advantage".  Using entangled photons as a quantum local oscillator (QLO), \citet{gottesman2012} showed the laser shot noise of conventionally heterodyne interferometry can be side-stepped, recovering similar sensitivity of today's ``direct detection'' methods but with potentially important practical advantages in the future. \citet{brown2022} recently reported first lab demonstration of quantum advantage using heralded quantum-entangled photons, and others have speculated on advantages for quantum-enhanced interferometry in a future world with quantum networks and quantum hard drives \cite[e.g.,][]{joss2021}.

\subsection{Software tools and resources}
\label{software}
A host of community tools exist to help interferometrists prepare observations, analyze data, make images, and search archives. The Jean-Marie Mariotti Center (http://jmmc.fr) is a clearinghouse of essential software resources including {\em searchCal} for finding calibrators, {\em ASPRO} for planning observations at any facility, {\em LITpro} for fitting simple models, {\em OImaging} for image reconstruction tools, {\em Bad Cal} for bad calibrator list, and the {\em OIDB} for archiving and searching for reduced data.  New python-based {\em PMOIRED} \citep{pmoired} also can fit a range of models as can julia-based {\em SQUEEZE, OITOOLS} (https://www.chara.gsu.edu/analysis-software/imaging-software) .  All these tools use the OI-FITS Standard for exchanging calibrated optical interferometry data \citep{oifits1,oifits2}.

\begin{summary}[INTERFEROMETRY PRIMER - SUMMARY POINTS]
\begin{enumerate}
\item A long-baseline interferometer is a Young's two-slit experiment, which measures Fourier components of the object brightness distributions.
\item Today's facilities have best angular resolution ranging from 0.5 to 5\,mas.
\item Atmosphere turbulence poses fundamental limitations that can be overcome using closure phases, adaptive optics, fringe-tracking, and dual-beam phase-referencing.
\end{enumerate}
\end{summary}

\section{ADVANCES IN INTERFEROMETRIC IMAGING} 
\label{imaging}

\begin{marginnote}[]
\entry{uv coverage}{Extent of the Fourier plane containing measured complex visibilities. The greater the uv coverage, the better quality images can be reconstructed.}
\end{marginnote}

The past decade has not only seen a breakthrough in sensitivity but also in imaging power.  It is striking that most of the results we highlight in \S\ref{breakthrough_astrophysics} are {\em images} and not {\em visibility curves}, a huge change from previous reviews of O/IR interferometry.  In this section we outline how this was made possible through effective use of 4- and 6-telescope arrays, new beam combiners that use all telescopes simultaneously, better beam combiner architectures to enhance calibration precision and data throughput, and also mature and proliferating image reconstruction software.

\subsection{Filling the uv-plane} 
\label{largearrays}

\begin{figure}[t]
\includegraphics[width=\columnwidth]{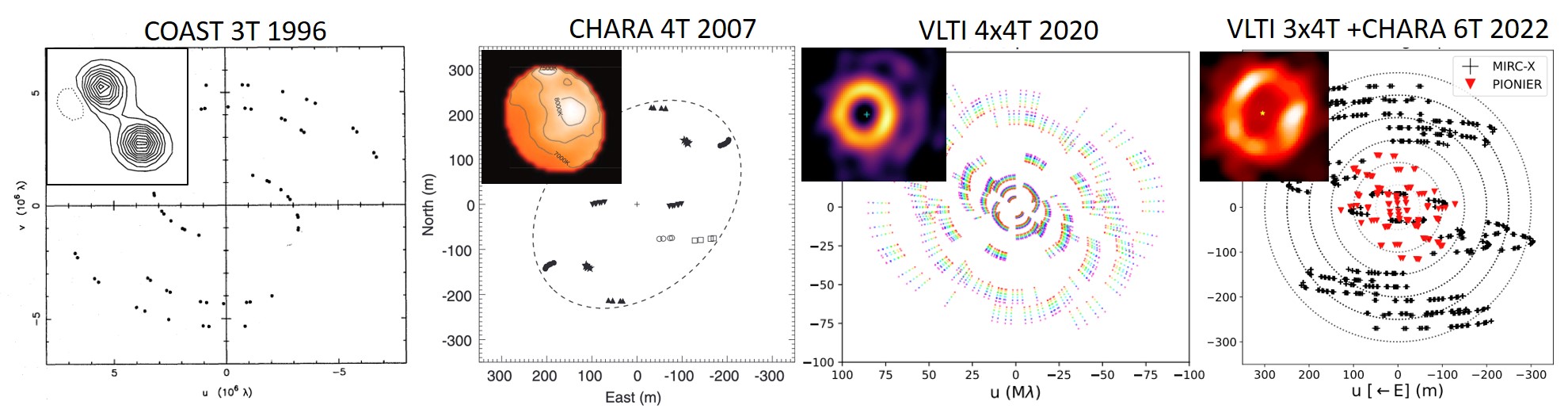}
\caption{The uv plane can now be densely filled using multiple configurations of four- and six-telescope arrays. The evolution over the past 20 years has been dramatic and we show here actual published uv coverage along with the reconstructed images\citep[][Ibrahim et al. 2022 submitted]{baldwin1996,monnier2007,kluska2020b}. 
}
\label{fig:uvcov}
\end{figure}

The O/IR interferometry field currently has three major imaging interferometers, NPOI (reconfigurable 6T), CHARA (fixed 6T), and VLTI-Auxiliary Telescopes (ATs, reconfigurable 4T) / VLTI-Unit telescopes (UTs, fixed 4T). This is reminiscent of the pre-ALMA age for millimeter-wave interferometers when many small arrays existed with $<10$ telescopes \citep{sargent1993}. The number of baselines in an $N$ telescope interferometer is $N(N-1)/2$ (\S\ref{closure}), thus even adding a few telescopes can substantially transform the imaging capabilities. Since each projected baseline is a single Fourier component of the true sky image, the imaging fidelity of an array is directly related to the ability to ``fill the uv-plane."  In addition to simply increasing the number of telescopes, we can improve uv coverage by a) using earth rotation to sample different baseline projections, b) employing movable telescopes at the same site and combining configurations (as long as object does not change in time), c) combining data from geographically different interferometers at nearly the same time, d) observing over a wide wavelength range to sample different resolutions when coupled with a wavelength-dependent model of source.   Figure\,\ref{fig:uvcov} shows the evolution of uv coverage over past decades.

Even when arrays were built with 4$+$ telescopes, beam combiners did not yet exist to combine all the beams at the same time. This was the case for NPOI, VLTI and CHARA which did not possess combiners that could measure all baselines and closure phases simultaneously until the past decade, often a decade after first light.  One part of the delay was caused by the cumbersome prospect of creating a 4+ beam combiner in bulk optics, which would take up large optical tables and require on-going tedious alignments. As already covered above, the classic ``pairwise" beam combiner works well for two or three telescopes, however, scales poorly for large number of telescopes.  Light from each of $N$ telescopes must be split $N-1$ ways and then combined with all other beams on $N \choose 2$ beamsplitters leading to $N (N-1)$ outputs. Further, bulk optics combiners historically have not produced as precise calibration as those using single-mode waveguides.  

\subsection{Better calibration and miniaturization with single-mode waveguides} 
\label{singlemode}

\begin{marginnote}[]
\entry{Integrated Optics (IO)}{Optical equivalent to electronic integrated circuits (ICs). Dielectric chips implanted with optical waveguides providing beam splitters, combiners, delay lines, nulling, and more}
\end{marginnote}

We already mentioned the breakthrough in calibration from single-mode fibers and evanescent wave fused couplers \citep[e.g., IOTA/FLUOR, ][]{foresto1998,perrin1998} in the 1990s, but this architecture did not lead to all-fiber beam combiners for imaging. This was due to the difficulties in constructing a network of fiber-based splitters and couplers while maintaining better than millimeter internal path lengths and controlling for differential dispersion and birefringence.  

\begin{figure}[t]
\includegraphics[width=\columnwidth]{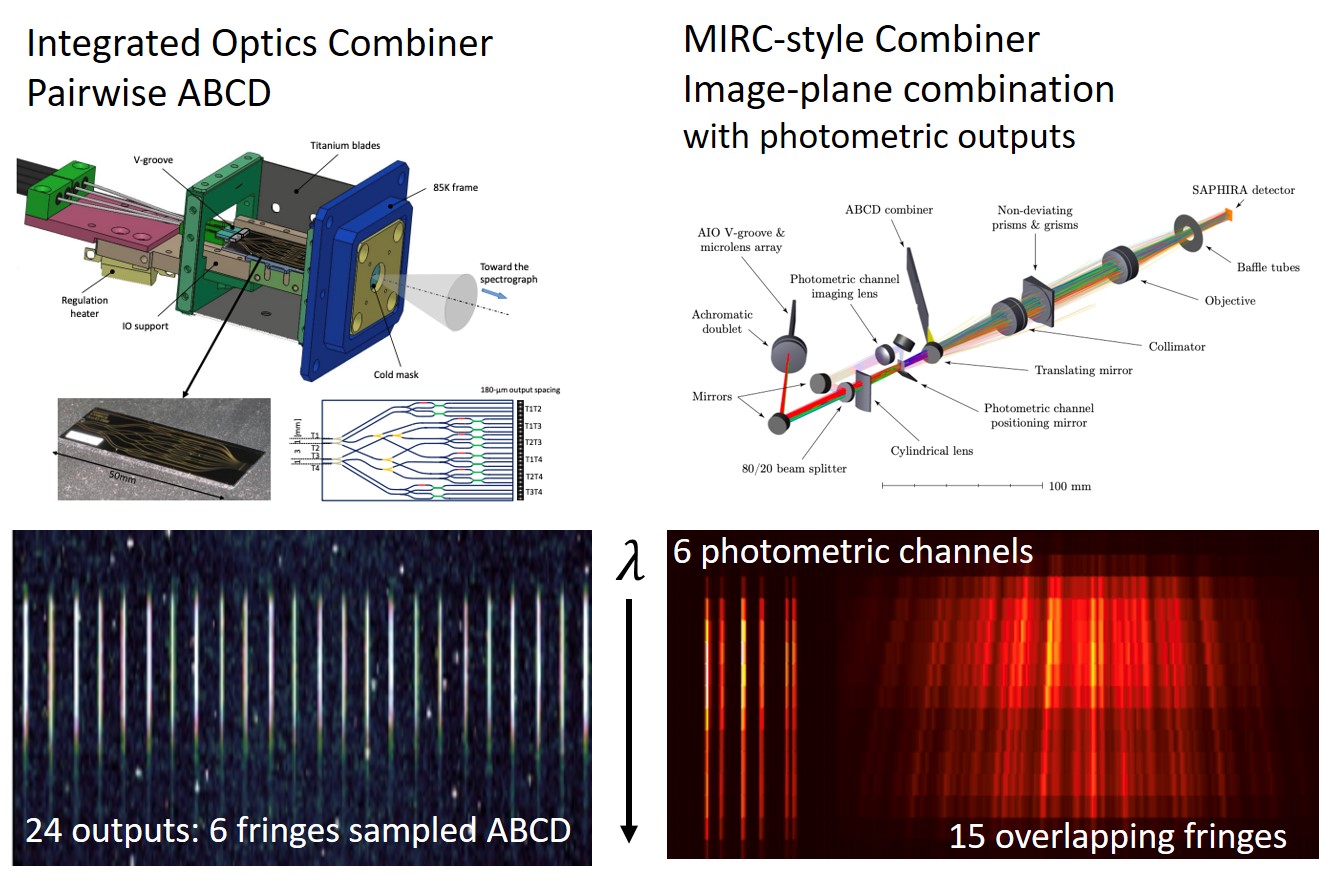}
\caption{The imaging breakthroughs of the past decade required a new generation of ``imaging" combiners. Today's imaging instruments have all adopted either a pairwise ABCD combiner using an IO \citep[left panel figures adapted from VLTI/GRAVITY instrument;][]{perraut2018} or a MIRC-style instrument using image-plane combination following single-mode fibers (right panel figures adapted from CHARA/MYSTIC instrument; Setterholm et al. 2022).
}
\label{fig:combiners}
\end{figure}

This engineering problem was first solved using planar waveguide circuits \citep{kern1997}, developed initially for telecommunications \citep[see recent review,][]{righini2014}.  In analogy to integrated circuits, integrated optics (IO) allows complex optical functions such as splitters, combiners, achromatized phase delays, and active phase modulation to be embedded in a thin dielectric slab using lithographic methods.  The first IO device used for astronomy was IOTA/IONIC \citep{berger2001} and this technology has since matured with a basic form where the input beams are split, recombined to form all interference pairs and with internal delays to sample four equally-spaced fringe phases.  The IO outputs are in a line that can then be dispersed on to an array detector. This so-called pairwise ABCD IO combiner is the basis for the H-band 4T combiner VLTI/PIONIER \citep{lebouquin2011}, the K-band 4T combiner VLTI/GRAVITY \citep{perraut2018}, the K-band 4T combiner CHARA/MYSTIC-ABCD, and the new H-band 6T combiner for the CHARA/SPICA fringe-tracker.  IO devices rarely can operate beyond about $\approx 30$\,\% spectral bandwidth (due to the requirement to have low bending losses while maintain high single-mode purity), thus each astronomical band (J,H,K) needs its own device.  Material absorption and scattering losses in the visible and thermal IR have limited IO use in these wavebands. Due to the need to minimize the curvature of waveguide bends, the chip length scales like $N^2$, incurring significant propagation losses ($\approx 50$\,\%) for current devices.  Better materials and new writing techniques continue to be developed in this active area, for instance see further discussion on nulling in \S\ref{nulling} and the recent advances in 3-D laser writing \citep[e.g.,][]{labadie2018}, which promises more compact combiners in more materials as the writing quality improves.

Another influential beam combiner architecture tailored for imaging with large N telescopes uses image-plane combination of light that has been spatially-filtered by fibers but with beam combination in the image-plane using conventional bulk optics. The CHARA/MIRC combiner pioneered this method, using conventional telecom fibers installed in a non-redundant linear array within a silicon v-groove mated with a microlens array. The interference fringes can then be focused onto a slit and dispersed, as for the IO devices. Bending losses tend to limit the usable bandwidth of IO devices, while single-mode fibers can have high transmission over $\approx 50$\,\% spectral bandwidths, for instance including $1.1-1.8\, \mu$m using standard telecom fibers or $1.5-2.4\,\mu$m using special low-OH silica or fluoride fibers. This style of combiner is limited to wavelengths where good fibers exist. The image-plane combiner is also susceptible to cross-talk \citep{mortimer2022}.  So-called MIRC-style combiners lie at the heart of current instruments MIRC-X \citep{anugu2020}, MYSTIC (Setterholm et al. 2022), SPICA \citep{mourard2018} and VISION \citep{garcia2016}. Figure\,\ref{fig:combiners} show schematics of both IO and MIRC-style combiners and how the dispersed interference patterns look in practice.

\subsection{Advanced image reconstruction algorithms} 
\label{imagereconstruction}

Many of the highest profile results in O/IR interferometry would not have been possible without advances in image reconstruction algorithms, beyond those developed by the radio community. Here we describe the current state-of-the-art but demur from detailed derivations or anything resembling a ``how-to".  We refer interested readers to the review of methods by \citep{baron2016, baron2020}, along with tutorial-like presentation by \citet{thiebaut}. Here you will find a brief overview of the current algorithms and future directions.

The radio community was first to attempt interferometric imaging and had initially been quite innovative in developing algorithms to account for the limited uv coverage of early facilities. The CLEAN \citep{hogbom1974} algorithm was tractable even with early computers and subsequently built into the standard radio packages AIPS and CASA. The basic principle behind CLEAN is to take a direct Fourier transform of the complex visibilities in the (mostly empty) uv plane, then to follow this with a deconvolution step to remove the large ``sidelobes'' that appear in the initial image. Since most of the artifacts come from the large {\em known} gaps in the uv-plane, this method works well for simple objects, especially when consisting of mainly point sources like binary or triple systems. Closure phases are incorporated using an iterative ``self-calibration'' scheme, applying positivity and limited field-of-view windows when carrying out the deconvolution \citep{readhead1978}.  Unfortunately,  CLEAN does a poor job with reconstructing smooth extended emission, does not intrinsically deal with error bars or closure phases, and smooths away details as a way to minimize artifacts. The flaws of CLEAN are most severe with small $N<10$ arrays like we have in O/IR interferometry.

\begin{marginnote}[]
\entry{CLEAN}{Traditional deconvolution technique for image reconstruction with interferometers}
\entry{Regularizer}{A mathematical metric that is optimized when fitting interferometric data with a model image}
\end{marginnote}

Faster computers now allow a more mathematically rigorous approach. The ``forward modeling'' approach considers all possible astronomical images as hypotheses and then choose the best set of images based on a combination of the best-fit to the observables ($\chi^2$) as well as any {\em prior} information, such as positivity, known field of view, etc. This is an example of an ill-posed inverse problem with no unique solution, so one must ``regularize'' the solution by introducing additional constraints. The first widely-adopted ``regularizer''  was the Maximum Entropy Method \citep[MEM;][]{gull1984} which attempts to maximize the so-called Entropy $S$ of the image: $S=- \sum_i f_i \log {\frac{f_i}{p_i}}$, where $f_i$ is the fraction of flux in pixel $i$ and $p_i$ is the Bayesian prior (often taken as constant). Conceptually we expect this to find the ``smoothest'' possible image consistent with the data, which is attractive for the astronomical utility.

The panoply of codes now available now often incorporate alternative regularizers to entropy, that go by a variety of names such  ``total variation'', ''uniform disk regularizer'', ``dark energy'', some of which prefer sharp edges while other select smooth edges. Thus the astronomer is now expected to choose the regularizer appropriate to their target \citep{renard2011}. Another way to approach image reconstruction is through ``compressed sensing'' where one tries to find a good-fitting image that has the fewest number of ``components'', where a component can be a something like a wavelet coefficient or derived from a dictionary of radiative transfer calculations.

Today's image reconstruction frontier involves regularizing across time to track moving blobs in dust shells and disks, regularizing across wavelengths to improve uv coverage but accounting for smooth changes in images with wavelength, and even imaging with a (latitude, longitude) pixel grid projected onto a rotating spheroid to better image spots on spinning stars. \citet{kluska2014} outlined a widely used approach where a central star with a known size and spectrum can be combined with a dust image which can have a different spectrum but which is otherwise ``grey.''  This SPARCO method is needed for imaging disks around young stars where the flux ratio between star and disk varies substantially across wavelength channels. 

Radio work on imaging with sparse uv coverage radically slowed in the 1990s, and the IAU Working Group on O/IR Interferometry began a series of meetings around 2000 to address the need for better imaging as the modern facilities were building up.  These meetings led to the OI-FITS data standard \citep{oifits1,oifits2} for calibrated optical interferometry data as no appropriate standard from radio existed.  In addition, a series of image reconstruction ``contests'' have been held during the SPIE astronomical instrumentation conference every two years  since 2004 to encourage development of advanced imaging methods and to recognize excellent contributions.  These double-blind contests typically used simulated data based on existing facilities and instruments with realistic noise \citep[e.g.,][]{beauty2004} or even were attempted on real data \citep{beauty2014}.  The result has been an innovative and dynamic subfield that has produced a range of community tools, usually publicly available: BSMEM \citep{buscher1994}, MACIM \citep{macim2006}, MiRA \citep{mira2008}, IRBis \citep{hofmann2014}, WISARD \citep{wisard2008}, SQUEEZE \citep{baron2016}, SURFing \citep{roettenbacher2016}, ROTIR \citep{rotir2021}, ORGANIC \citep{organic2020}, $\rm G^R$ \citep{gravity_gr}.  This work has now fed-back to radio interferometry, informing the imaging approach of the Event Horizon Telescope (EHT) project \citep[e.g.,][]{lu2014}.

\begin{summary}[ADVANCES IN INTERFEROMETRIC IMAGING - SUMMARY POINTS]
\begin{enumerate}
\item Interferometric imaging requires as many telescopes as possible to fill the uv-plane.
\item Within the past decade the major facilities have been able to combine all their available telescopes, with single-mode optics at the core of today's beam combiners.
\item Specialized image reconstruction algorithms have been developed for O/IR interferometry, advancing the field over earlier work by radio astronomers.
\end{enumerate}
\end{summary}

\section{BREAKTHROUGH IN SENSITIVITY} 
\label{breakthrough_sensitivity}

Initially, astronomical interferometers were made from small aperture telescopes or siderostats with a diameter less than a few times the atmospheric Fried parameter $r_0$.  

The sensitivity advantage of large telescopes for interferometry has been recognized early on, and both the VLT \citep{lena1979,beckers1990} and Keck telescopes \citep{colavita1998,colavita2013} have been designed and implemented to also operate as an interferometer. The use of AO then increases $r_0 \approx 0.1 ... 1\mbox{ m} \rightarrow D \approx 10\mbox{ m}$, i.e. the ``coherent volume" $r_0^2 (c \tau_0)$ (\S\ref{atmosphere}) increases by factor $\gg 100$, such that fringes of moderately faint objects $m_K \approx 10...11$\, mag can be detected within the atmospheric coherence time $\tau_0 \approx \mbox{\,few ms}$. This allows for stabilizing the fringes, and thereby opens up a factor 1000 - 100000 longer exposures times $\rightarrow T_{exp} \approx$\,minutes -- hours, and reaching out for yet another factor 1000 fainter objects $m_K \approx 19$\,mag by dual-beam interferometry. 

\begin{figure}[t]
\includegraphics[width=\columnwidth]{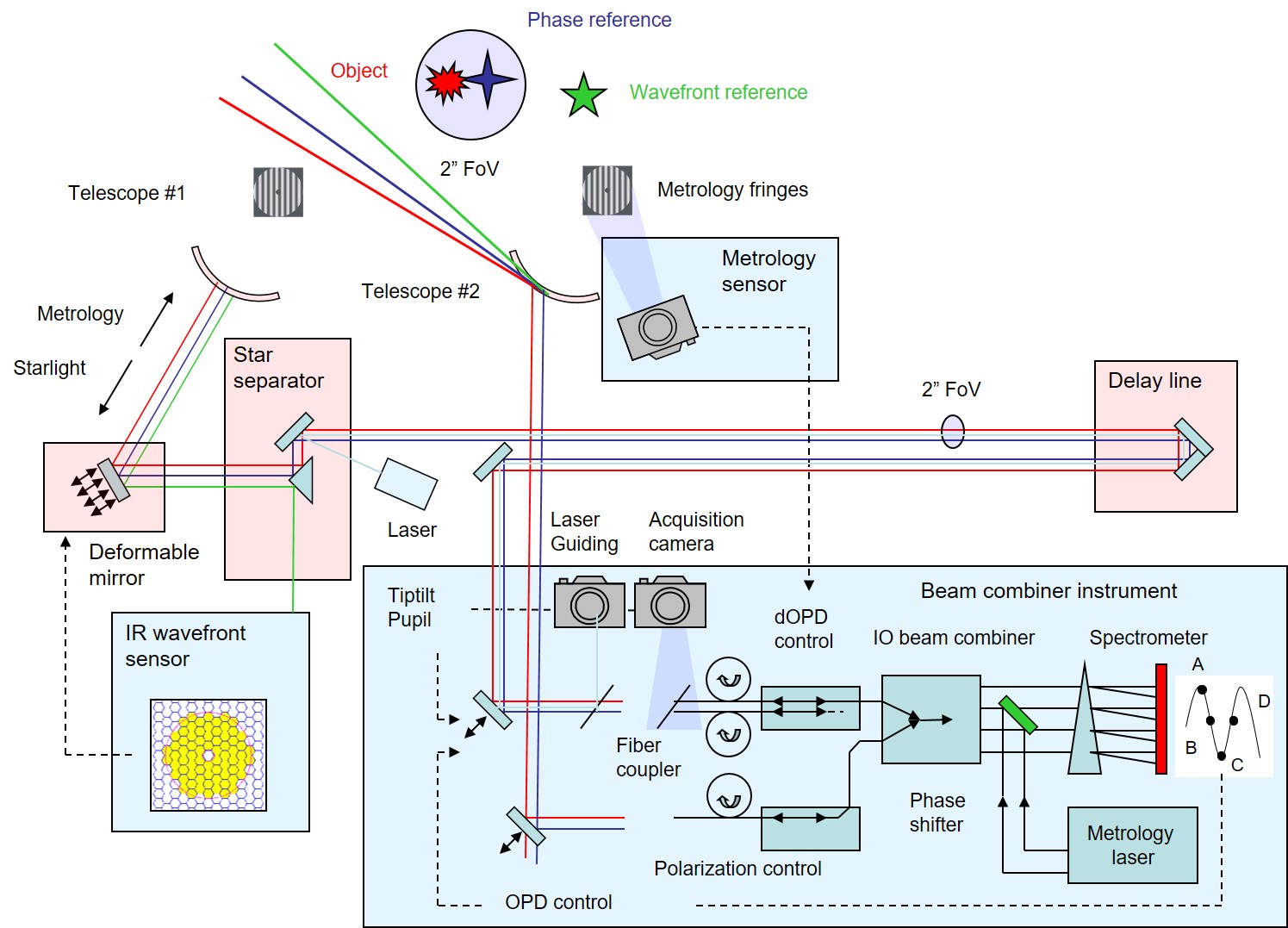}
\vspace{0 cm}
\caption{ Overview and working principle of a modern, large telescope interferometer at the example of GRAVITY: each telescope is equipped with AO to provide a diffraction limited beam, which is then transported through mirrored delay lines to the beam combiner instrument. The instrument provides two beam combiners, one for fringe-tracking, the other optimized for long exposure, high spectral resolution interferometry of the science target. Off-axis fringe-tracking is done by either separating the phase-reference and science target at the instrument level (narrow field of view) or at the telescope level (wide field of view). In the latter case the two beams are propagated separately to the beam combiner instrument. The optical path length within the observatory is controlled via several laser metrology systems, delay lines and differential delay lines. The beams are stabilized through active field and pupil tracking, again taking advantage of laser metrology. For clarity, we show only two telescopes, and one AO, delay line, and beam combiner \citep{gravity_first_light2017}. }
\label{fig:modern_interferometer}
\end{figure}

However, the technical hurdles were substantial, and it was only GRAVITY that brought all techniques together for routine science operation. Figure\,\ref{fig:modern_interferometer} illustrates the complexity and major subsystems of such a modern interferometer.

\subsection{Sensitivity gain from large telescopes} 
\label{largetelescopes}

\begin{figure}[t]
\includegraphics[width=16cm]{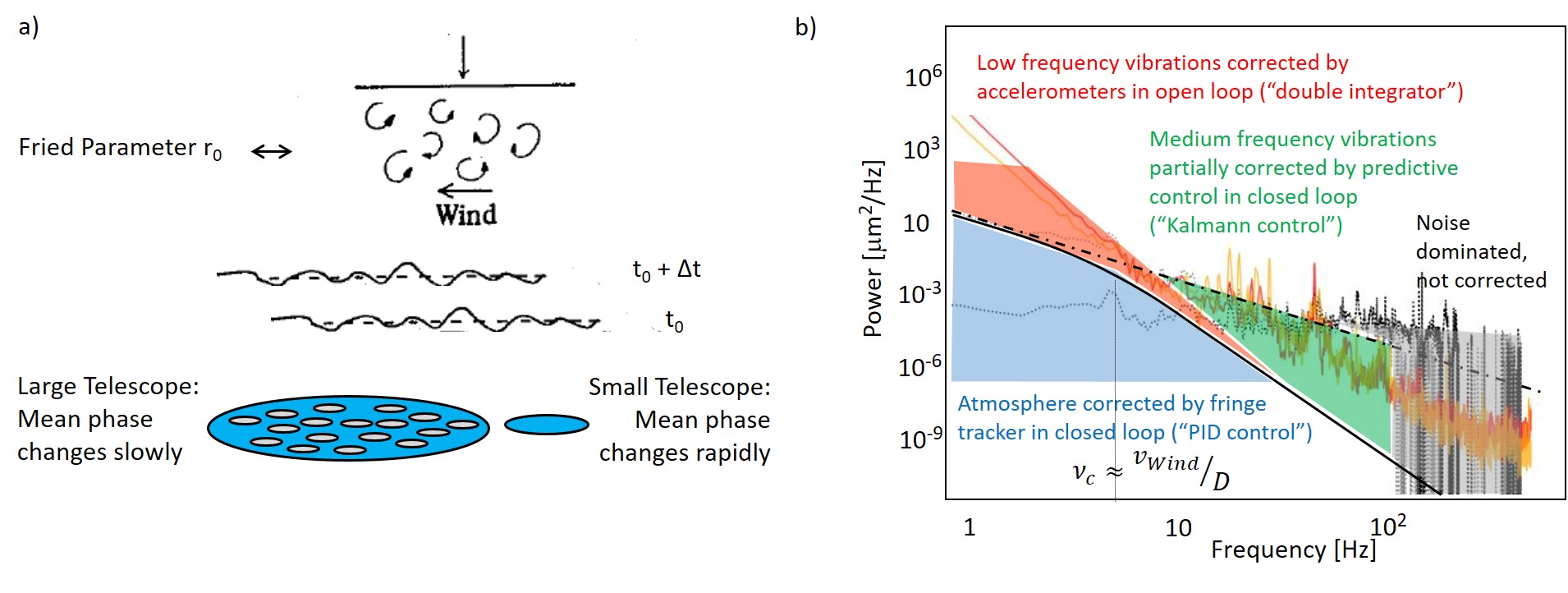}
\vspace{-0.5cm}
\caption{Interferometry with large telescopes. a) large apertures average out the small scale atmospheric perturbations, thereby increase the coherence time as compared to small telescope. The resulting temporal power spectrum (b) of the mean phase (black thick curve) exhibits a break around a typical frequency of $\approx \frac{v}{D}$, where $v$ is the wind speed, and D the telescope diameter (adapted from \cite{conan1995}). Vibrations (red/yellow lines) exceed the atmospheric path length fluctuations for large telescopes. They are corrected by a combination of accelerometers (open loop control) and fringe-tracking (closed loop) for low and intermediate frequencies up to few ten Hz. Vibrations at higher frequencies - especially from optics not monitored by accelerometers - can be partially corrected by predictive control with the fringe-tracker, but ultimately limit the current fringe-tracking residuals (black).}
\label{fig:largetelescopes}
\end{figure}

\begin{marginnote}[]
\entry{Coherent volume}{Combination of Fried parameter and coherence time $r_0^2 (c \tau_0)$ indicating sensitivity of an interferometer}
\end{marginnote}

On top of collecting area, there are more fundamental advantages of large telescopes for interferometry. They arise from phase averaging effects across the aperture, which increase the characteristic timescales for fringe-tracking, and from a smaller diffraction limit, thereby enabling the observations of crowded regions. In the following we quantify each respective advantage, which taken together yield a large telescope advantage $\propto D^{2 ... 5}$.

The Signal-To-Noise (SNR) ratio in observations of point sources increases with the telescope diameter $\propto D^2$ for background limited, single-mode interferometric observations \citep{beckers1990}. The reason is that the thermal background for diffraction limited observations is irrespective of telescope size (\'etendue = collecting area $\times$ receiving solid angle = $\lambda^2$), while the signal increases $\propto D^2$, i.e. by factor 100 when going from one to ten meter telescopes. At the same time, the telescope primary beam shrinks in area with telescope diameter $\propto D^{-2}$. In the crowding limit, the number of objects in the observed field of view is reduced by a factor $\propto D^2$, and therefore the SNR per object increases accordingly $\propto D^2$, i.e. by another factor 100 when going from one to ten meter telescopes.  

Taken together, the SNR increase for crowding limited observations in the IR is $\propto D^4$. This holds for interferometry as well as for diffraction limited observations with single telescopes. In O/IR interferometry with its comparably poor uv coverage, the smaller field of view can be even decisive for the observation of complex scenes. For the Galactic Center, e.g., the stellar density is about 70 stars brighter than 19.5 mag within the diffraction limited field of view of an 1 m diameter telescope, too many for the sparse uv coverage of a four telescope interferometer. It is the factor 100 smaller diffraction area of 10-m telescopes, which reduces the complexity to a few stars per pointing and thereby allows for the deep imaging and the accurate astrometry presented in \S\ref{galacticcenter}. 

\begin{marginnote}
\entry{Outer scale of turbulence $L_0$}{Distance, at which the phase structure function falls significantly below the Kolmogorov power spectrum}
\entry{Cutoff frequency}{Frequency, above which the power spectrum falls significantly below the extrapolation from lower frequency}
\end{marginnote}

Next, the coherence time increases with telescope diameter, because single-mode interferometers are probing the average phase of the pupil, thereby smoothing out the temporal fluctuations (Figure\,\ref{fig:largetelescopes}a). The increase in coherence time is $\propto D^{1/6}$ for moderate size telescope $r_0 \ll D \ll L_0$ \citep{kellerer2007}, and $\propto D^{7/6}$ when D approaches the outer scale of turbulence $L_0$. However, current telescopes ($D \lesssim  10$\,m) are smaller than $L_0$ (typically few 10\,m for good astronomical sites \citep{ziad2016}), and take only moderate advantage of the increased coherence time. The more relevant advantage comes from the shape of the phase power spectrum (Figure\,\ref{fig:largetelescopes}b): the averaging effect leads to the suppression of especially the high frequency perturbation, and results in a turnover of the atmospheric Kolmogorov power spectrum at a frequency $\approx 0.3 \frac{v}{D}$ \citep{conan1995}, where $v$ is the wind speed. This cutoff frequency will reflect in the required control bandwidth of the fringe-tracker, which in turn is ultimately set by the detector integration time. In the case of readnoise-limited short exposures, the SNR per exposure will therefore increase with the exposure time $\propto \frac{D}{v}$. In practice, the gain will be somewhere in between the above two regimes, i.e. $\propto D^{1/6 ... 1}$

While the sensitivity boost SNR\,$\propto D^{2 ... 4}$ has already materialized with GRAVITY, it has not yet been able to take advantage of the increased coherence and fringe-tracking exposure times $\propto D^{1/6 ... 1}$ because of excessive vibrations of large telescopes (\S\ref{fringetracking}). 

\subsection{Adaptive optics} 
\label{adaptiveoptics}

AO is the technique to correct the wavefront distortion from the Earth's turbulent atmosphere and to deliver diffraction limited images for large telescopes. A wavefront sensor measures the distorted wavefront of a bright star, which is then compensated in real-time with a deformable mirror, so that also the image from nearby objects will be sharpened. The original concept goes back to \citet{babcock1953} and was followed up independently and in parallel in military and astronomical applications. See e.g. \citet{beckers1993} and \citet{davies2012} for reviews of its history, principle and applications in astronomy.

AO is also the prerequisite to efficiently interfere the light from telescopes with a diameter larger than the Fried parameter. Its decisive role was recognized already in the proposed implementation of the VLT interferometer \citep{beckers1990}: while the sensitivity (SNR$^2$) - the faintest object detectable for a given observing time - of broad-band interferometric observations scales only moderately $\propto D^{1/3}$ with the diameter $D$ of large telescopes in the case of seeing-limited (so-called multi-speckle) observations, the sensitivity increases $\propto D^4$ for diffraction-limited, single-mode interferometric observations. Therefore dedicated AO was implemented from the very beginning in the 8-m UTs for VLTI \citep{arsenault2003} and later upgraded with the GRAVITY IR AO \citep{hippler2020}. The Keck interferometer took advantage of the multi-purpose Keck AO \citep{wiziinowich2000}. The smaller 1.8m diameter ESO ATs and the 1.0m CHARA telescopes are now also equipped with AO \citep{woillez2019, brummelaar2012}. 

\begin{marginnote}
\entry{Adaptive optics (AO)}{Technique to correct image blur from turbulent atmosphere and recover diffraction limited resolution of telescopes}
\entry{Laser guide star (LGS)}{Artificial star created with a laser, used for AO when no bright natural star is close to the observed object}
\end{marginnote}

While the limiting magnitude of small telescope interferometers is set by fringe-tracking, large telescopes interferometers are limited by AO. The reason is that larger telescopes require ever higher order AO, because each atmosphere turbulence cell has to be corrected independently, and therefore the AO limiting magnitude is independent of the size of the telescope. Fringe-tracking, however, only measures and stabilizes a single (average) phase irrespective of telescope size, and the limiting magnitude thus increases with telescope diameter $D$. Because of additional light losses between the AO at the telescopes and the fringe-tracker in the interferometric laboratory, the limiting magnitude curves are shifted relative to each other, and the transition between the two regimes typically arises at 1--2 m telescope diameter as used in CHARA and the VLTI ATs. 

Laser guide star (LGS) AO overcomes this limitation \citep[e.g.,][]{beckers1993, davies2012}. Here the high order wavefront correction is done on an artificial star, created by a laser, and projected high up in the stratosphere. In this case the limiting magnitude of the AO is set by the need for a tip/tilt reference star (two degrees of freedom per telescope), which is comparable to the fringe-tracking limit (one degree of freedom per telescope, but reduced optical throughput). A first demonstration of LGS AO assisted interferometry was done at the Keck interferometer \citep{colavita2013} shortly before its shutdown. The GRAVITY+ project \citep{eisenhauer2019} currently upgrades all VLT 8m telescopes with LGS and new AO (\S\ref{enhancing_sensitivity}).

\subsection{Fringe tracking and vibration control for minute long exposures} 
\label{fringetracking}

\begin{marginnote}
\entry{Fringe tracking}{Technique to stabilize the interferogram of two telescopes for long exposures}
\entry{Kalman filter}{Recursive algorithm, which estimates the internal state of a system from a series of noisy measurements.}
\entry{Closed-loop bandwidth}{Maximum frequency up to which a control system can damp perturbations by a factor 2 (3dB) or better}
\end{marginnote}

Fringe-tracking is the technique to track and correct the phase delay between telescopes which is introduced by the turbulent atmosphere. Among the first to propose and later implement an active tracking and stabilization of the fringe position using a servo-loop were \cite{shao1977,shao1980}. A number of instruments followed the same path and implemented fringe-tracking to allow the interferometric observation of faint sources by actively stabilizing on a nearby bright source \citep{colavita2003,gai2003,delplancke2008,kok2013}.

The first element of a fringe-tracker is the phase sensor, a beam combiner which measures the current fringe position (\S\ref{singlemode}). A controller then compares the measured fringe position with a target position and commands an actuator (typically a fast piezo-mounted mirror) to correct the position. The different control-loop states (e.g. phase delay tracking close to the optical path zero, group delay tracking if the fringe is off by $>\lambda$ or fringe-search in case the fringe has been lost) are handled by a state machine.  

An important limitation of fringe-tracking are vibrations from the telescopes, instruments, and infrastructure. They excite the mirrors along the optical train and result in optical path length fluctuations with pronounced peaks at the frequencies of the excitation and at the mechanical resonance frequencies of the mirror and mounts. The detailed power spectrum (Figure\,\ref{fig:largetelescopes}b) of the optical path variation is complex, with isolated peaks, forests of unresolved peaks, and a pseudo-continuum following a red power-law. 

At frequencies below a few Hertz (Hz), the optical path fluctuations are completely dominated by vibrations, with an amplitude of order several to ten $\mu$m. At the VLTI and LBTI, e.g., they are measured by accelerometers placed on the primary, secondary and tertiary mirror, and corrected by piezo-driven mirrors in the main delay lines in feed-forward open loop \citep[e.g.][]{dilieto2008,boehm2017}. After the bulk of these vibrations have been corrected, the turbulence of the Earth's atmosphere dominates the optical path length fluctuations up to few ten Hertz. This is the primary domain of the fringe-tracker. At higher frequencies, the atmospheric perturbations have only little power, and on top, the averaging effect from large telescope aperture sets in (Figure\,\ref{fig:largetelescopes}a), damping the high turbulence frequencies even more. 

For large telescopes, however, this high frequency range is again dominated by vibrations excited by, e.g., ventilators (around 48 Hz) and closed-cycle coolers of cryogenic instruments (around 80 Hz). The vibrations often come in the form of a forests of multiple, nearby frequencies, e.g. inductance motors with a slightly different frequency slip (typically few percent). With fringe-tracker frame rates of 100-500\,Hz and corresponding closed-loop bandwidth of a few ten Hz, most of these vibrations are beyond the control bandwidth. The vibrations are therefore corrected with predictive control, very much like in AO \citep[e.g.][]{guesalaga2012,kulcsar2012}. One possibility are phase-locked oscillators to track and correct a small number of vibration peaks \citep{dilieto2008,colavita2010}. An alternative approach is to compensate the vibration spectrum and the atmospheric turbulence at the same time. 

This can be achieved with a \cite{kalman1960} controller, a predictive algorithm whose commands are based on a model of the identified disturbance components. \cite{menu2012} successfully ported the Kalman filtering to four-baseline fringe-tracking, which was subsequently implemented in the GRAVITY fringe-tracker by \cite{choquet2014} and \cite{lacour2019}. The Kalman filtering outperforms the classical PID control, reducing the (vibration-driven) fringe-tracking residual error on the VLTI UTs from $500-1000\rm\,nm\,rms$ to $\sim250\rm\,nm\,rms$ under typical conditions. On the small ATs, the fringe-tracking residuals reaches residuals $<100\rm\,nm\,rms$ \citep{lacour2019}. Only the successful compensation of the telescope vibrations enabled minute long integration times in dual-field interferometry.

\subsection{Dual-beam interferometry} 
\label{dualbeam}

\begin{marginnote}
\entry{Dual-beam interferometry}{Simultaneous interferometry of two objects. Allows observing of faint objects by fringe tracking on a bright, nearby reference star. Also used for astrometry and phase-referencing for improved aperture synthesis}
\entry{Astrometry}{Position measurement, here between two objects}
\entry{Phase-referencing}{Technique, which optically links the fringe phase of two objects by an internal metrology.}
\end{marginnote}

Dual-beam interferometry refers to the simultaneous interferometric observation of two widely separated objects ($\theta \gg$ coherent field-of-view) contained inside the atmospheric turbulence isopistonic patch (\S\ref{atmosphere}). The technique was first described by \citet{shao1992} and later implemented by \citet{colavita1999} at the Palomar Testbed Interferometer using star separators located at the focus of the telescope, which feed two independent beams to separate interferometric instruments. In the early 90s the interest in dual-field interferometry was primarily driven by the promise of high precision astrometry (\S\ref{astrometry}) and its application in the detection of exoplanets through the reflex motion of their host stars (in preparation for space astrometry missions for exoplanets). The possibility to fringe-track on a bright object (\S\ref{fringetracking}) and to stabilize and observe fringes of a much fainter nearby target was tentatively explored by \cite{lane2003}, but was not pushed beyond the limiting magnitude of the interferometer around $m_K = 5$\,mag.

The dual-ﬁeld instrument PRIMA \citep{delplancke2008} was designed to equip the VLTI with astrometric and phase-referencing capabilities. The main science interest in exoplanet detection put an emphasis on astrometry rather than pushing the sensitivity of the interferometer. Delays in the project and competition from GAIA \citep{gaia2016} led to a stop of the astrometric efforts, while the dual field capability of the infrastructure was briefly explored. The ﬁrst dual-ﬁeld phase-referenced observations, which reached a magnitude of $m_K = 12.5$ pushing the sensitivity by a factor 10 compared to direct observations, where carried out by the ASTRA instrument \citep{woillez2014}. The emerging scientiﬁc capability of the Keck Interferometer was stopped from flourishing by the shut-down of the facility in 2012. It took another 5 years to unlock the full potential of phase-referenced imaging and to push the sensitivity limits by a factor 1000 compared to direct detection. The GRAVITY instrument \citep{gravity_first_light2017} routinely oﬀers mas resolution imaging for objects fainter than $m_K = 19$\,mag (\S\ref{galacticcenter}). 

\subsection{Sub-electron readnoise infrared detector arrays} 
\label{detectors}

Irrespective of detailed implementation, the tracking and compensation of atmospheric perturbations with AO or fringe-tracking requires exposure times shorter than the atmospheric coherence time. At typical frame rates of 1\,kHz, i.e. exposure times of 1\,ms, even bright objects are photon starved. This means that the performance of the sensor is limited by the readout noise of the detector \citep[e.g.][]{finger2016}. At the same time, the 1\,kHz frame rate of typically a few thousand pixels requires a comparably high analog bandwidth of order MHz for reading the detector pixels. 

The early IR fringe-tracking systems (e.g. PTI fringe-tracker, FINITO at VLTI, FATCAT at Keck) used NICMOS3, PICNIC or HAWAII arrays with typical readout noise for double correlated sampling of $\approx20-30\,$e$^-$ with exceptional systems reaching $\approx 5-10\,$e$^-$ \citep{millangabet1999,colavita1999,gautam2003}. Although the IR detectors grew enormously in size from 0.065\,Mpixel to 16\,Mpixel over the last three decades, the improvements in terms of read noise have been marginal.

\begin{marginnote}
\entry{Infrared detectors}{Detectors made from semiconductors with a smaller bandgap than Silicon (used for optical detectors), thereby sensitive to the lower energy IR photons}
\entry{Complementary metal–oxide– semiconductor (CMOS) detector}{Detector array technology, which separates the photodiodes from the readout electronics (other then CCD detectors, which combine detection and readout in the pixels), utilized in IR detectors}
\entry{Avalanche Photo Diode (APD)}{Photodiode detector which amplifies the signal by a cascade of photo-electron multiplications, the semiconductor equivalent of a photomultiplier tube}
\entry{Double correlated sampling}{Reading the voltage of the pixel in the beginning and end of an exposure to remove fixed pattern noise in CMOS detectors}
\entry{Quantum efficiency}{Percentage of photons converted to photoelectrons}
\end{marginnote}

In order to overcome the CMOS noise barrier, ESO started in 2007 a program together with SELEX (now LEONARDO) to develop HgCdTe based electron avalanche photodiode arrays (eAPD) for the NIR \citep{finger2010}. An independent eAPD development under the name "RAPID" was started by SOFRADIR, CEA-LETI and a consortium of research institutes \citep{feautrier2014}. HgCdTe is a well suited detector material for avalanche multiplication since the mass of the electron is much smaller than the mass of the holes and the APD process results in pure electron multiplication. HgCdTe is a direct semiconductor, i.e. electron-hole pairs are created without phonon interaction. This means that large avalanche gains with almost no excess noise are possible with HgCdTe based APDs \citep{finger2016}. After several development cycles of solid state engineering, the eAPD arrays have matured and resulted in the 320$\times$256 pixels SAPHIRA arrays. The first units were implemented in the GRAVITY fringe-tracker and Coud\'{e} Infrared AO units (CIAO). At short integration times ($\sim1$\,ms) the SAPHIRA arrays achieve subelectron noise ($< 0.2$\,e$^-$), high APD gain (up to 700), and an excess noise of only $\approx 1.3$ \citep{finger2016,finger2017}. The introduction of the SAPHIRA arrays in AO and interferometry and the corresponding noise reduction from $\sim10\,$e$^-$ to $<1\,$e$^-$ has been revolutionary, and has played a considerable role in the success of VLTI/GRAVITY. A number of interferometer and AO facilities at observatories (e.g. Keck, Subaru, Kitt Peak, CHARA and Palomar) also adopted the SAPHIRA \citep[e.g.][]{goebel2018}.

The latest detector generation Mark14, grown by metal organic vapour phase epitaxy (MOVPE), has extended the sensitivity range to $\rm 0.8-2.5\,\mu m$ \citep{finger2016}. Current developments include high-speed $512\times512$ pixel eAPD array for AO applications on 30-40\,m telescopes (Finger et al. 2019, 2022, in press), and large format science detectors (Claveau et al. 2022, Feautrier et al. 2022, both in press). 

\begin{summary}[BREAKTHROUGH IN SENSITIVITY - SUMMARY POINTS ]
\begin{enumerate}
\item Sensitivity of O/IR interferometers scales with high powers of telescope size.
\item Adaptive optics, vibration control, and fringe-tracking allow for minute long exposures -- a factor 1000 longer than the atmospheric coherence time.
\item Dual beam interferometry enables routine milli-arcsecond resolution imaging of objects fainter 19 magnitude.
\item eAPD detectors revolutionised the field of adaptive optics and interferometry.
\end{enumerate}
\end{summary}

\section{PRECISION INTERFEROMETRY} 

\subsection{Narrow angle astrometry} 
\label{astrometry}

Ground-based astrometry is fundamentally limited by Earth’s atmosphere. This limitation manifests itself in scintillation, image blurring and image motion in case of single dish imaging. In case of multiple apertures, astrometry is limited by the fringe jitter introduced by the atmospheric turbulence. The jitter of two stars however is correlated, depending on their separation and the aperture size (or baseline in case of an interferometer). With narrow-angle astrometry we denote the regime when the angular separation between two sources is small enough that their beams experience essentially the same atmospheric perturbations. 

The narrow angle regime is defined by \cite{shao1992} as the angular separation $\theta \ll B/h$, with the baseline $B$ and the effective turbulence height $h$. The advantage of narrow-angle astrometry is that the atmospheric jitter of two sources is partially correlated and cancels in a differential measurement. Among the first to recognize the potential of narrow-angle astrometry was \cite{lindegren1980}. The application in interferometry was later proposed by \cite{shao1992}. They realized the possibility of $\mu$as astrometry with long baseline interferometry. Using SCIDAR measurements of the atmospheric turbulence height profile at Mauna Kea, \citet{shao1992} derived an expression of the differential astrometric error for $\theta \ll B/3000$\,m. The residual error behaves as $\sigma\approx 1.45\cdot10^{-3}\,B^{-2/3}\,\theta \,t^{-1/2}\, \rm [arcsec]$, with the source separation $\theta$ in arcseconds and the integration time $t$ in seconds. Of particular importance in the narrow angle regime is the $\sigma \propto B^{-2/3}$ dependence, which explains why large interferometer arrays are key to high-precision astrometry. Assuming a baseline of 100\,m and a separation of $2\,\rm arcsec$, the residual atmospheric error averages to $\rm \sim10\,\mu as$ within 5\,min of integration time.

\begin{marginnote}
\entry{Narrow angle regime}{Angular separation $\theta << B/h$, with the baseline $B$ and the effective turbulence height $h$, where the atmospheric tilt jitter of two objects is correlated.}
\entry{SCIDAR}{Scintillation detection and ranging  technique, to measure the atmospheric turbulence profile.}
\end{marginnote}

The principle of differential astrometry between two objects on sky relies on the measurement of the optical path difference $\rm\Delta OPD$ of the two objects, which originates from the geometric delay introduced by the different projection of the baseline $\vec{B}$. The basic astrometric equation ${\rm \Delta OPD}=\vec{B}\cdot \vec{s}+\Delta \rm OPD_{int}$ relates the measured $\Delta \rm OPD$ between the two objects, their differential optical path inside the interferometer $\Delta \rm OPD_{int}$ and the object separation $\vec{s}=(\vec{\alpha}-\vec{\beta})$, where $\vec{\alpha}, \vec{\beta}$ are the positions of the two objects. A simple sensitivity analysis illustrates the required accuracies: $\delta s=\frac{\delta B}{B}\cdot s + \frac{\delta \Delta \rm OPD}{B}$. Let's assume the desired astrometric accuracy is $10\,\rm\mu arcsec$, the baseline is $B=100\,\rm m$ and the source separation is $s=1\,\rm arcsec$: this means that the baseline needs to be calibrated to millimeter accuracy, and the combined optical path delay error must be measured to few nm accuracy. The astrometric equation becomes significantly more complex when systematic uncertainties due to the residual atmosphere, baseline calibration, instrument and telescope alignment as well as metrology, polarisation and dispersion errors are considered. The following list gives an overview of major astrometric error sources:
a) Baseline errors originate from the difficulty to measure and calibrate the $B_{\rm NAB}$, which is physically realised by the end points of the metrology (\S\ref{baselines}). Depending on the location of the end points, $B_{\rm NAB}$ can be impacted by differential telescope flexure or runout.  
b) OPD errors (${\rm \Delta OPD_{S}}=\phi_{FT}\cdot \frac{\lambda_{FT}}{2 \pi} - \phi_{SC}\cdot \frac{\lambda_{SC}}{2 \pi}$) introduced by phase measurement and wavelength calibration errors in the fringe tracking or science channel. 
c) Metrology errors (${\rm OPD}_{M}=\phi_M\cdot \frac{\lambda_{M}}{2\pi}$), here most importantly the wavelength, because the metrology traces differential OPDs of the order few 10-100\,mm. 
d) Dispersion errors $\Delta L_{\rm fib,air}(\frac{n_{\rm fib,air} (M)}{n_{\rm fib,air} (S)}-1)$ from inaccurate calibration of the different refractive index between the metrology and the science wavelength in air or glass. 
e) Pupil mis-registration of the science and metrology beams in combination with pointing errors. They lead to phase errors of the order $\Delta \vec{\alpha}\cdot (\vec{P_{S}}-\vec{P_{M}})$. 
f) Polarisation induced astrometric errors from differential birefringence in the interferometer arms. Since the metrology laser is linearly polarised, birefringence can introduce an effective path difference between the laser and the unpolarised science objects. 

These and more astrometric errors have been studied by \cite{colavita2009} and \cite{woillez2013} as well as in particular for the GRAVITY instrument by \cite{lacour2014,lacour2014b}. The impact of polarisation has been studied by \cite{lazareff2014}. 

\subsection{Interferometric baselines revised}
\label{baselines}
For infinitely small telescopes, the interferometric baseline is the separation vector between the apertures. It gives the optical path difference between the two telescopes when observing a point-like object (\S\ref{interferometry}), sets the spatial frequency in interferometric imaging and the van-Cittert Zernike relation (\S\ref{vanCittertZernike}), and results in the differential optical path difference between two neighboring objects in phase-reference observations (\S\ref{fringetracking}). For large telescopes, however, the baselines for the three cases are not identical, and each of them follows a different physical realization. The effects are often non-intuitive, especially for narrow-angle astrometry (\S\ref{astrometry}). \cite{woillez2013} and \cite{lacour2014} concisely reviewed the three concepts in preparation for the GRAVITY instrument, building on the learning from the earlier ASTRA \citep{woillez2010} and PRIMA \citep{delplancke2008} experiments. 

\begin{marginnote}[]
\entry{Baseline B and $\vec{B}$}{Separation and vector (3D) between a pair of telescopes}
\entry{Projected baseline}{Separation between telescopes as seen from the observed target (2D)}
\entry{Wide angle baseline}{Separation between pivot points of telescopes - delay equation}
\entry{Imaging baseline}{Separation between telescope pupils, van Cittert - Zernike relation}
\entry{Narrow angle baseline}{Separation between metrology end points - dual-beam astrometry}
\end{marginnote}

\begin{figure}[t]
\includegraphics[width=11.5cm]{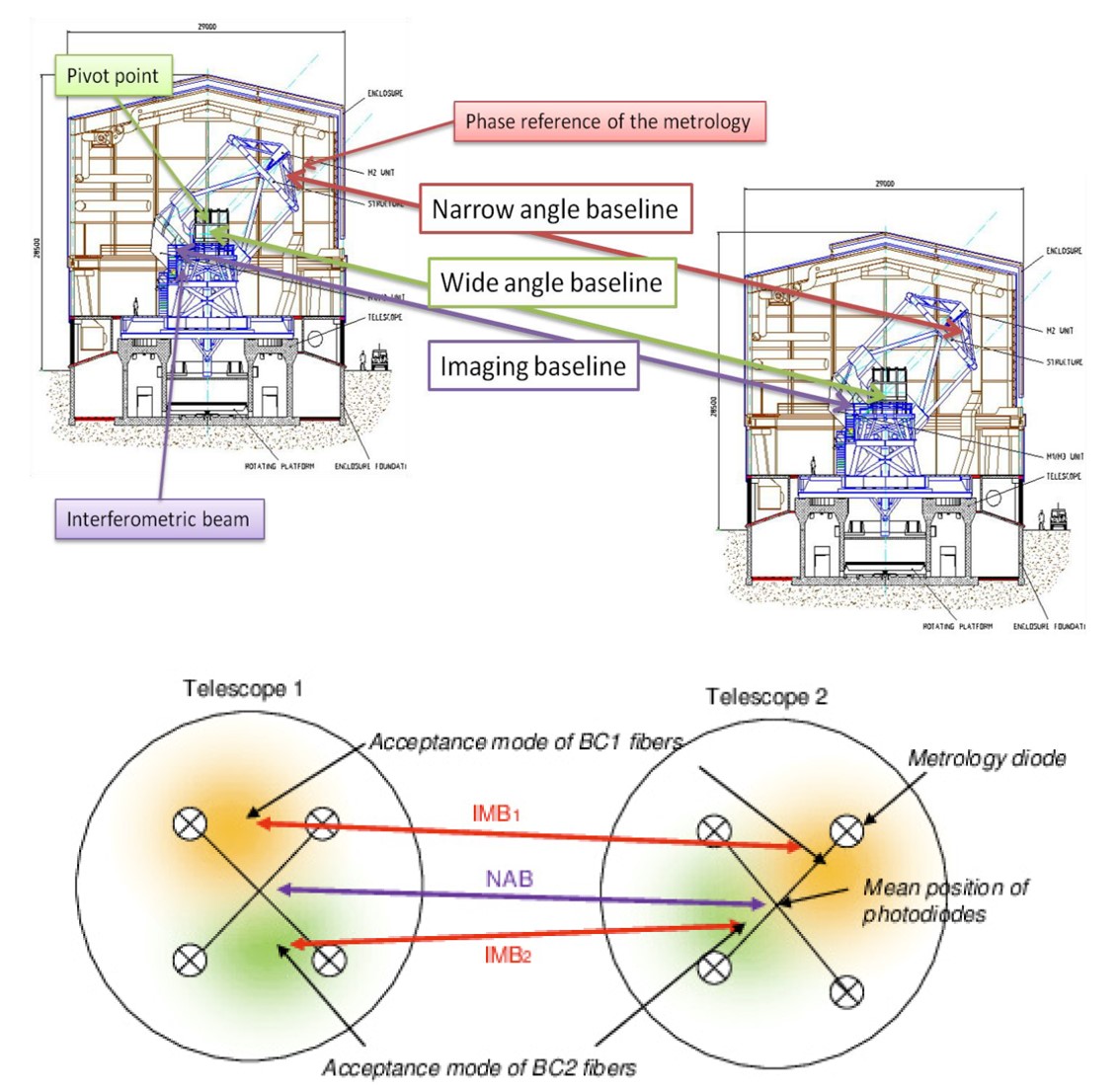}
\caption{Baselines in O/IR interferometry: the separation between the telescopes of an interferometer is called baseline. In detail, we distinguish three types, each of them with a specific property and physical realization: the wide angle baseline (WAB) obeys the delay equation, and is realized by the pivot points of the telescopes. The imaging baseline (IMB) sets the spatial frequency in interferometric imaging, and is realized by the entrance pupil weighted by the Gaussian beam of the instrument single-mode fibers. For a dual-beam interferometer, the imaging baselines can be different for the two objects because of optical misalignment. The narrow angle baseline (NAB) is the separation between the endpoint of metrology system, which measured the internal path difference between two objects (adapted from \citep{lacour2014}).}
\label{fig:baselines}
\end{figure}

The wide angle baseline is the separation between the pivot points of the telescopes. Since each telescope points towards the object, there is no phase gradient across the aperture, and all light rays have the same optical path length irrespective of their pupil position. The wide angle baseline tells where to preset the delay lines to find the fringes, and when precisely known, allows for measuring the absolute position of the star on the celestial sphere. The wide angle baseline is calibrated from stars with known coordinates, and for precision wide angle astrometry, the position of the telescopes pivot points are measured and monitored with dedicated laser metrology, e.g., at NPOI \citep{npoi1988}. 

The imaging baseline sets the spatial frequency sampled by the interferometer. The observed complex visibility is then the Fourier transform of the objects intensity distribution (\S\ref{vanCittertZernike}). The imaging baseline is given by the autocorrelation of the telescopes apertures as seen from the object \citep{thompson2017}. More accurately, it includes the coupling of the electric field probed by the beam combiner instrument -- i.e. the Gaussian mode of the instrument's single-mode fiber -- and the beam relay from the telescope to the beam combiner. 

Finally, the narrow angle baseline links the distance of the interferograms of two separated objects, each measured with its own beam combiner. The differential optical path between the two interferograms is measured by laser metrology, probing the path length from each beam combiner to the telescopes (\S\ref{astrometry}). More accurately, the narrow angle baseline is the separation between the end-points of the laser metrology, where the path length difference is measured, again, as seen from the object. The metrology endpoint for narrow angle astrometry are best placed in front of the telescope, e.g., for GRAVITY on the spider holding the secondary mirror \citep{lippa2016}, because even small wobbles from the optical relay would alter the apparent position of internal end-points and thereby corrupt the narrow angle astrometry.

\subsection{Laser metrology in interferometer} 
\label{metrology}

A key component of any dual-beam interferometers is a dedicated metrology system, which provides an optical link of the two beam combiners and their science interferograms. The metrology monitors the internal path length with nm accuracy to enable high precision astrometry. In some cases, the system also serves as a feedback for differential delay lines, which compensate the optical path difference of the two objects. As discussed in \S\ref{baselines}, the end points of the metrology, i.e. the points to which the internal optical paths are measured, define the narrow angle baseline $B_{\rm NAB}$ of the interferometer. Several metrology concepts have been developed for dual-field interferometers. The metrology systems implemented in PTI \citep{colavita1999} and PRIMA \citep{leveque2003} rely on two heterodyne Michelson interferometers, which measure for each object the optical path change between the telescopes. The internal path difference between the objects $\Delta \rm OPD_{int}$, corresponds to the difference between the path variations recorded by the two heterodyne interferometers.     

\begin{marginnote}
\entry{Lock-in amplifier}{A type of amplifier that can extract a signal with a known carrier wave frequency from an extremely noisy environment.}
\end{marginnote}

The PRIMA metrology launches the lasers at the center of the pupil plane close to the beam combiners, and they are retro-reflected at the telescopes. The internal path difference $\Delta \rm OPD_{int}={(T1-T2)}_{FT} - {(T1-T2)}_{SC}$, i.e. the difference between the path variations, is recorded by photodiodes. The signal is filtered and the individual heterodyne signals are mixed such that the disturbance to be monitored is directly coded in the phase of the carrier signal. The GRAVITY metrology \citep{lippa2016} splits the laser light into three beams with fixed phase relations. Two of the beams are faint and injected backwards into the two beam combiner chips. The third, high-power beam is overlaid on top of the two faint ones after they have passed the fibers within the instrument, and acts as an phase-preserving amplifier for the detection. This approach minimizes the inelastic scattering of the metrology light in the instrument fibers. At each telescope, the three beams interfere in the pupil plane and form a fringe pattern. By temporal phase modulation of the two faint beams at different kHz frequencies the phase signal of the two light paths can be extracted using of lock-in amplifiers. The metrology measures the path variation between the objects to a single telescope. The difference of the variation between two telescopes provides the desired internal $\Delta \rm OPD_{int}={(FT-SC)}_{T1} - {(FT-SC)}_{T2}$ between the two objects.

A subtle but important distinction between the two metrology concepts originates from the way how the $\Delta \rm OPD_{int}$ is measured. The PRIMA-like design allows to measure and correct vibrations occurring between the telescopes. However, a serious caveat of the PRIMA concept is that the metrology cannot trace the beam before the AO deformable mirror (DM), because it would imprint its deformation in the $(T1-T2)$ path, thereby folding the atmospheric perturbations to the metrology signal. As a consequence, the metrology end points need to be downstream of the DM far away from the telescope primary mirror, which penalizes the $B_{\rm NAB}$ stability. In the GRAVITY concept, the two laser beams $(FT-SC)$ go through the same DM, i.e. the difference signal does not see the deformation, which allows extending the metrology up to the telescope pupil, a decisive advantage for the stability of $B_{\rm NAB}$. This comes at the cost that vibrations impacting the fringe-tracker and science beam, are not traced by the metrology.   

Both metrology concepts have in common that they rely on frequency stabilised lasers, which operate at wavelengths shorter than the science band to minimize stray light. In both cases, the laser frequency $\nu=c / \lambda$ needs to be sufficiently stable such that the corresponding OPD error is less than 1\,nm over a differential internal OPD of a few -- 100 mm. This corresponds to a frequency stability of better than $10^{-7...-8}$. 

\subsection{Field- and pupil stabilization}
\label{beamstabilization}

The use of single-mode fibers mandates the accurate control of the star's position to better than the diffraction limit. This is typically achieved by a combination of AO (\S\ref{adaptiveoptics}) and/or guider at the telescope, and a lab- or instrument provided secondary guiding on the object itself to correct the image motion from imperfect beam relay. This two stage approach provides the large field of view at the telescope for picking a bright, nearby star for telescope guiding, so that the secondary guider can run with longer exposure times and accordingly higher sensitivity. The lab guider \citep[e.g.][]{gitton2004,crawford2006} is typically a dedicated camera, operating at a different wavelength and serving several instruments. This comes with the disadvantage of non-common guiding errors, e.g., from atmospheric refraction. The field stabilization requirements are especially hard for narrow angle astrometry (\S\ref{astrometry}), for which even small tilt errors (order 10 mas) reduce the astrometric accuracy through cross-terms with pupil errors. In this case the guiding is directly done on the fiber by slightly (radius few mas) modulating the tip/tilt at high frequency (few 10 Hz) and using the correlation between the measured flux and modulation signal (in analogy to a lockin amplifier) as feedback to the control loop \citep[e.g.][]{bonnet2006}. Some beam combiner instruments also provide internal wavefront sensors \citep{anugu2018} for secondary guiding of low order aberrations, in particular focus. 

\begin{marginnote}
\entry{Telescope guider}{Camera to track and stabilize the telescope on bright star}
\entry{Secondary guiding}{Provides extra correction of perturbations between telescope and instrument}
\entry{Variable curvature mirror}{A mirror for which the radius of curvature can be actively adjusted, e.g. by air pressure}
\end{marginnote}

It is also the narrow angle astrometry which requires accurate pupil control, typically to well below a percent of the telescope diameter. Because of the difficulty to see and track the pupil on faint objects, pupil guiding is implemented by means of laser beacons, either taking advantage of the path length metrology \citep[e.g.,][]{woillez2014}, or dedicated laser beacons launched from the spider arms holding the telescope's secondary mirror \citep{anugu2018}. Most interferometers do not re-image the telescope pupil, but transfer only a collimated beam to beam combiner instruments, and therefore have zero field of view. The VLTI is the only interferometer with variable curvature mirrors \citep{ferrari2003} in the retro-reflectors of its delay lines to dynamically transfer the telescope pupil and to provide a coherent field of view of $2-6$\,'' for the beam combiner. As the delay lines move to compensate the pathlength difference between the telescopes, the curvature of this mirror is adjusted either blindly, or in closed loop on the laser beacons launched at the telescope \citep{anugu2018}.

\subsection{The role of optical aberrations in interferometry} 
\label{abberations}

The introduction and implementation of single-mode fibers \citep{froehly1981,shaklan1987,foresto1998} and resulting improvements in the calibration was a breakthrough in O/IR interferometry. The spatial filtering from single-mode fibers converts phase errors across the telescope pupils in amplitude fluctuations, which can be monitored and then corrected in the calculation of the visibility (\S\ref{singlemode}). It thus removes the random phase fluctuations of the turbulent atmosphere as the dominant calibration error \citep[e.g.,][for a first demonstration]{perrin1998}. Dynamic aberrations like atmospheric tip/tilt result in an increased field of view \citep{perrin2019}. Static optical aberrations have a more subtle effect on the interferometric measurement. The phase of the electric field from a point source is then not constant anymore across the Airy peak of the telescope point spread function, but is a function of the position within the field of view. 

\begin{marginnote}
\entry{Optical aberrations}{Deviation from perfect optical image formation, typically described as wavefront error in the pupil plane, but also imprints on the phase of the electric field in the image plane}
\end{marginnote}

In radio-interferometry, the effect is known as direction-dependent (complex) gains \citep{bhatnagar2008}. For O/IR interferometry, it was for the first time described and corrected in \cite{gravity_aberration}. Even for diffraction limited optics the static aberrations are of order $\lambda/10$, translating to comparable signatures in the interferometric measurements, or when expressed in position errors on sky, 1/10 the interferometric beam size for objects located at the edge of the interferometers field of view. In GRAVITY, e.g., the calibration of the field dependent phase errors is done by scanning the field of view with the instrument internal calibration unit. The correction is then applied via a modified van Cittert-Zernike theorem in the forward modelling as part of the model-fitting (e.g., when fitting the measured visibilities and closure phases with a binary star) or image-reconstruction \citep{gravity_gr}. 

\subsection{Nulling} 
\label{nulling}

Perhaps the technique with potential to deliver the most extreme precision is {\em nulling interferometry,} with a goal of $\ll 10^{6}$ contrast to detect Earth-like planets in the MIR. Akin to phase-mask coronagraphy on single-apertures \citep[see, e.g., review by][]{mawet2012}, nulling uses destructive interference to remove host-star light while letting light from a faint nearby object to be detected. Nulling interferometry potentially has the combination of angular resolution and high contrast to characterize a large number of Earth-like planets around other stars, especially in the thermal IR where many molecular biomarkers can be seen in the planetary atmospheres \citep[e.g.,][]{quanz2022}.  

\begin{marginnote}
\entry{Nulling}{Using destructive interference to cancel the light from a point source in one of the beam combiner outputs, thereby reducing the photon noise}
\entry{Coronagraphy}{Technique to blacken out the direct light from a star and to suppress its diffraction rings.}
\entry{(Exo-)Zodiacal light}{Glow of diffuse starlight scattered on interplanetary dust (in the solar system and around exoplanets}
\end{marginnote}

The principle behind nulling was laid out by \citet{bracewell1978} and \citet{angel1997}: if one interferes light from two telescopes onto a beamsplitter with a  $\pi$ relative phase shift, then the electric fields destructively interfere and one beamsplitter output is completely dark, except the light from a second object (e.g. exoplanet) slightly offset on sky. In practice though there are many difficulties, including chromaticity of phase delay, electric field amplitude and polarization matching, wavefront aberrations, stellar diameter leakage, and fringe-tracking stability.  The first sky nulling was performed by \cite{hinz1998} using sub-apertures from a single telescope, where a $\pi$ phase shift was created using glass combinations, and achieved $\approx 4$\,\% null depths at best in open loop. Long-baseline nulling at the Keck Interferometer \citep[][using a field inversion for nulling]{serabyn2012} and the LBTI measured the exozodiacal dust contributions around nearby stars: \citet{colavita2009b} reported raw null depths of $1.5-2$\,\% in the IR and \citet{defrere2016} also achieved  $\approx 1$\,\% null depths at the LBTI. These surveys \citep{millangabet2011,defrere2016} concluded that exozodiacal dust is generally not large enough to interfere with future exoplanet searches. Due to the importance for nulling to the future of exoplanet studies, theoretical and experimental work remains active.  We point the reader to recent explorations of the optimal nulling architectures \citep[e.g.,][]{guyon2013,hansen2022}, ways to combiner nulling with closure phases \citep{lacour2014b}, nulling within IO \citep[e.g.,][]{hsiao2010,errmann2015,martinod2021}, and development of better IR-friendly materials for IO, such as lithium niobate \citep{hsiao2009} and chalcogenides \citep{kenchington2017}.  There is a new exoplanet-focused instrument VLTI/HI-5 \citep{defrere2018} under construction that will push nulling down to L-band (3.5$\mu$m) for the first time, and there is an active consortium proposing a space nulling interferometer, the Large Interferometer for Exoplanets \citep[LIFE;][]{quanz2022}. More on nascent space interferometer efforts in \S\ref{space}. Space is indeed the place for nulling since ground-based interferometry is tragically limited by atmospheric turbulence and thermal emission that degrade the achievable contrast far from fundamental limits.

\begin{summary}[PRECISION INTERFEROMETRY - SUMMARY POINTS ]
\begin{enumerate}
\item Interferometers provide much more accurate astrometry than single telescopes.
\item It is necessary to distinguish between three kind of baselines: wide angle baseline, imaging baseline, and narrow angle astrometry baseline.
\item Laser metrology measures path length differences to nanometer accuracy.
\item Active field- and pupil control is key to cancel second order astrometric error terms.
\item Aberrations introduce phase variations across the field of view and limit the astrometric and imaging accuracy, but can be corrected by forward modelling. 
\item Nulling is the interferometry equivalent to coronagraphy to block out the star light.
\end{enumerate}
\end{summary}

\section{FUTURE DIRECTIONS} 

So far we have concentrated our discussion of the breakthrough in sensitivity on the gain from large telescopes, AO, fringe-tracking, and dual-beam interferometry. But where do we stand in comparison to fundamental limits - quantum noise and background noise? How much to gain, and which directions to go?

\subsection{Far from fundamental limits} 
\label{fundamentallimits}

The quantum limit for the measurement of the phase from a two telescope interference is given by the Heisenberg uncertainty principle $\Delta \phi \Delta N < \frac{1}{2}$, where $\Delta N = \sqrt{N}$ is the uncertainty in the number of photons N in the measurement \citep[e.g.,][]{townes2000}. The resulting SNR of the fringe contrast in the photon noise limit is given by SNR\,$= V \sqrt{N}$. The number of photons needed to detect fringes (SNR\,$= 5$) from a partially resolved object ($V$\,=\,0.5) is $\approx 100\ \mbox{photons}$, which for the case of two 8-m diameter telescopes and observing at K-band translates to 31\,mag for an one hour exposure. This is about 100,000 fainter than currently reached with GRAVITY. How far are we from the atmospheric atmospheric and thermal background limit? In this case, the noise is dominated by the background, and the SNR of the fringe detection is SNR\,$= V \frac{N}{\sqrt{N+B}} \approx V \frac{N}{\sqrt{B}}$, where B is the number of background photons. For a single-mode instrument operating at the diffraction limit, the atmospheric background at K-band is $\approx 10^3$\,photons/s. The thermal emission from a blackbody with the average temperature of the laboratory of 16\,°C translates to $\approx 10^4$\,photons/s, roughly a factor 10 more than the atmospheric background. The number of photons needed for the detection of a partially resolved object in one hour is then $\approx$\,few\,$10^4$, corresponding to $\approx 25$\,mag, or factor few hundred fainter than currently possible. 

\begin{marginnote}[]
\entry{Quantum limit}{Resulting from the Heisenberg un- certainty relation, that energy (number of photons) and arrival time (phase) cannot be simultaneously measured to better than the Planck constant}
\entry{Background limit}{Photon noise from thermal emission of the atmosphere, telescopes and warm optics}
\end{marginnote}

\subsection{Enhancing sensitivity} 
\label{enhancing_sensitivity}

Why are current O/IR interferometers falling several orders of magnitudes behind the fundamental limit? It is throughput, detector noise, instrumental background, and coherence loss. Here we illustrate the situation on the example of GRAVITY, and sketch out possible improvements from the ongoing upgrade to GRAVITY+ \citep{eisenhauer2019}.

The worst offender is throughput: the total detective QE is $\approx 0.1-1$\,\%. The losses are dominated by AO performance (Strehl ratio typically $10-40$\,\%), the beamtrain from telescopes to instrument (throughput T $\approx 35$\,\%), IO (T $\approx 54$\,\%),  and "traditional optics losses" (e.g., grism efficiency $\approx 25-50$\,\%). On top there is a factor 2 loss from splitting the light between fringe-tracker and science beam combiner, and a factor 3 from the pairwise combination of the 4T array. Some losses can be reduced with current technology: taken together, better LGS AO, improvements in traditional optics, and going to dual-beam interferometry, will bring a factor $10-20$ better throughput, thereby increasing the detective QE up to $\approx 10$\,\%. The losses from the pairwise beam combination remains a fundamental issue, because there is no noise-free amplification at O/IR wavelengths. Focal plane beam combination \citep[e.g., the so-called hyper-telescope;][]{labeyrie1996} can partly overcome this limitation, but face practical limitation from detector noise and difficult calibration. 

\begin{marginnote}[]
\entry{Detective quantum efficiency}{Product of optical transmission and detector quantum efficiency, including all losses}
\entry{Worst offender}{Term dominating sensitivity or error budget}
\end{marginnote}

The second worst offender is noise: for minute-long exposures, GRAVITY observations are limited by the laser background from the metrology. Reducing the laser power or the interleaved operation will bring the laser background to below the detector noise, improving the sensitivity by factor $1.5-5$. At that stage ``zero-noise" eAPD detectors (\S\ref{detectors}) can again revolutionize sensitivity. Their long-exposure noise is currently dominated by a high dark current, but revised diode structures, readout circuits and optimized operation have already led to a demonstration of dark current 0.025\,e$^-$s$^{-1}$pix$^{-1}$ \citep{atkinson2018} and readnoise $\lesssim 3$\,e$^-$ in minute long exposures (\citeauthor{finger2019} 2019, 2022, in press). Also large format eAPD detectors are in the coming (Claveau et al. 2022, in press, Feautrier et al. 2022, in press). These detectors are then expected to reduce the noise of high-spectral resolution observations by another factor $2-10$.

The last offender is coherence loss: while polarization mismatch between telescopes is typically well corrected, fringe-tracking residuals are catastrophic ($\gtrsim 500$\,nm with frequent fringe losses) for observations at the sensitivity limit, when the fringe tracker control bandwidth is reduced. For large telescopes, the coherence loss is then dominated by vibrations (\S\ref{fringetracking}). Expanding accelerometers to the full coud\'{e} optical train (Bigioli et al. 2022, in press), together with further reducing the source of vibrations \citep{woillez2018}, should allow for 100 nm fringe-tracking also for 8-m telescopes (e.g. for exoplanet work), as well as increase the coherence at the current sensitivity limit by factor 2 (e.g., extragalactic objects). 

All together, higher throughput, reduced instrument background, zero-noise detectors, and better vibration control, will enhance the sensitivity of current interferometers by another $1-2$ orders of magnitudes, with a limiting magnitude then beyond $m_K \approx 22$\,mag.

\subsection{Enhancing sky coverage} 
\label{skycoverage}

GRAVITY's leaps in sensitivity by factor 1000s resulted from dual-beam interferometry and  fringe-tracking on a nearby reference star. Because it splits the light of the two objects in the instrument, the maximum separation is 2\," for the UTs. GRAVITY Wide \citep{gravity_wide} has extended its dual-beam capability to also take advantage of the previously installed PRIMA dual-beam infrastructure. The maximum separation between the two objects can then be up to several 10\,", limited only by the Earth's turbulent atmosphere. The limit is set by the isopistonic angle $\theta_P$, at which the phase between the two objects become uncorrelated. 

\begin{marginnote}[]
\entry{Sky coverage}{Fraction of sky area / objects in a certain direction, which can be observed with a given instrument}
\end{marginnote}

The sky coverage is ultimately given by the combination of the isopistonic angle and the limiting magnitude for fringe-tracking. Both increase with telescope diameter, and we will show below that the advantage of large telescopes for dual-beam interferometry even exceeds the SNR advantage outlined in \S\ref{largetelescopes}. First, the isopistonic angle increases with telescope diameter $\theta_P \propto D^{1/6} (const. + D)$ \citep{elhalkouj2008, boskri2021}, because in this case the beams from the two stars partially overlap, and thereby see the same atmospheric perturbations (Figure\,\ref{fig:largetelescopes}). As a result, the isopistonic area for off-axis interferometry around a fringe-tracking star increases $ \propto D^{1/3 ... 7/3}$. For typical atmospheric conditions, the isopistonic angle increases from $\approx 12$\," to $\approx 50$\," when going from 1-m to 10-m telescopes. Second, the gain in SNR $\propto D^2$ and the reduced bandwidth requirement for fringe-tracking $\propto D^{1/6 ... 1}$ relax the brightness requirement by a factor $\propto D^{-2.2 ... -3}$, which for a star luminosity function $N({\rm Stars~brighter~given~flux~S}) \propto S^{-3/2}$ increases the sky density of suitable stars $\propto D^{3.3 ... 4.5}$.

\begin{figure}[t]
\includegraphics[width=12cm]{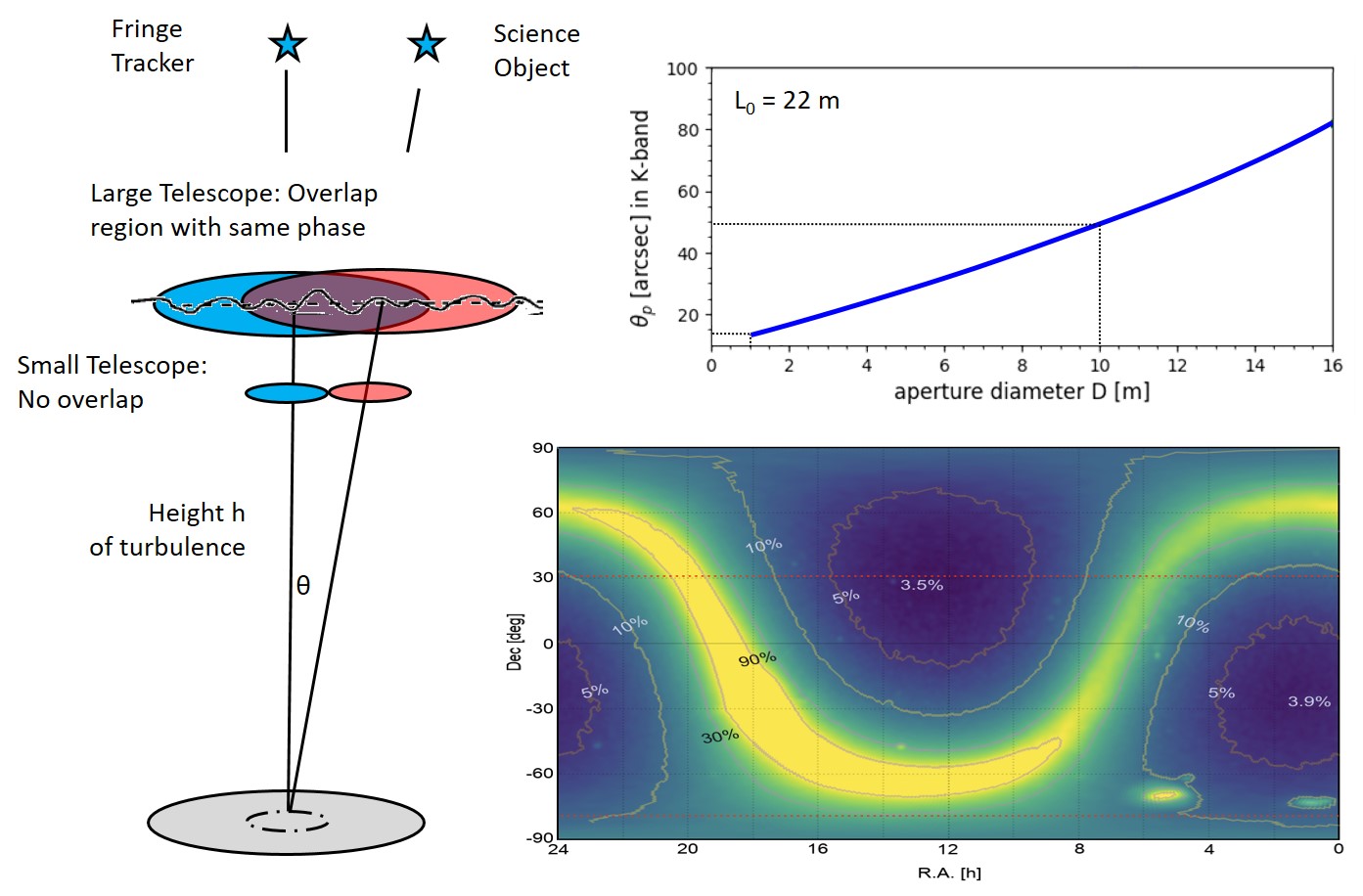}
\caption{Enhanced sky coverage for large telescopes. Left: when observing two objects simultaneously, the two beams overlap for large telescopes, thereby increasing the isopistonic angle with telescope diameter (top right, following \citet{elhalkouj2008}). Bottom right: The resulting sky coverage for 8m telescopes with a fringe-tracking limiting magnitude $m_K = 13$\,mag }
\label{fig:skycoverage}
\end{figure}

Taken together, the larger isopistonic area and the access to fainter fringe-tracking stars increases the sky coverage for dual-beam interferometry $\propto D^{3.6 ... 6.8}$. Vibrations and imperfect AO will eat up some of the advantages of large apertures, but it is only the 10-m class telescopes, in combination with LGS AO, to provide full (in the galactic plane) and substantial (for extragalactic observations outside the galactic plane) sky-coverage for off-axis fringe-tracking (Figure\,\ref{fig:skycoverage}).

\subsection{Enhancing wavelength and baseline dynamic ranges}

\begin{marginnote}[]
\entry{Dynamic range}{Maximum achievable contrast. In interferometric imaging driven by number of baseline, in model-fitting by signal-to-noise ratio}
\end{marginnote}

The wavelength range covered by O/IR interferometry is continuing to grow. MATISSE \citep{lopez2022} extended VLTI imaging capabilities to observe the L, M, N bands (3-12$\mu$m) simultaneously with 4 telescopes. Using the GRAVITY fringe-tracker will push limiting magnitudes especially for observations with higher spectral resolution, although only a few results are out as of this date. VLTI will also stretch to shorter wavelengths with the J-band BIFROST instrument, under construction \citep{bifrost2022}. Visible imaging with 0.3 mas angular resolution should now be possible at NPOI and CHARA, and we expect results in the near future for dwarf stars, interacting binaries, gaseous disks, and stellar surfaces.  Visible imaging will be boosted as NPOI integrates new 1m telescopes with AO into the system. Improved imaging needs better uv coverage and this means adding telescopes and baselines. While harder, there is progress here too as CHARA is adding a 7th mobile telescope that will be connected to the array with single-mode fibers, leading to potentially longer baselines as well as more dense uv coverage.  Ultimately, new facilities are needed to make a breakthrough in imaging and the MROI hopes to do this with 10 movable telescopes. While MROI will not have substantially longer baselines than CHARA, the uv-plane will be radically better sampled with virtually no gaps in the uv-plane, making high fidelity imaging truly possible.  This project is partially funded with first fringes expected in 2023.  The Planet Formation Imager project \citep[PFI;][]{pfi_spie2018} explored the science potential of 12x3m array with 1.2km baselines and found an exciting science case for detecting giant exoplanets as they are forming in young disks using K and L band nulling, but also concluded that sensitivity was insufficient to go below a few Jupiter-masses.  As the SNR for observations of point-sources scale $\sqrt{N_{\rm tel}} D^2$ in the background-limit (\S\ref{largetelescopes}),  truly revolutionary capabilities await ground-based arrays of ten or more 8-m class telescopes (i.e, a scaled up VLTI-UT) rather than many small telescopes -- or a move to space where the IR background is smaller by millions or even billions.

\subsection{Space interferometry} 
\label{space}

\begin{marginnote}[]
\entry{Formation flying}{Maintaining relative separation and orientation between multiple spacecrafts - prerequisite for large baseline interferometry in space}
\end{marginnote}

O/IR interferometry is well-suited to space, as all the major difficulties from the ground are missing: atmospheric turbulence that limit coherent integration times, light losses from complex optics to transport and delay light as stars move across sky, and severe thermal IR backgrounds. The first attempts in 1990s and 2000s to build space interferometers were ultimately not pursued, namely the Space Interferometer Mission (SIM, aimed at precision astrometry) and two MIR nulling interferometers -- DARWIN (ESA) and the Terrestrial Planet Finder Interferometer (TPFI) --  designed to detect and characterize Earth-like planets. A new flagship mission called LIFE, the Large Interferometer for Exoplanets \citep[e.g,.][]{quanz2022} is being proposed. The nulling interferometer with 4x2m telescopes working in the thermal-IR aims at detecting 550 exoplanets, with 25-45 being rocky planets in their habitable zones (HZ). For 3.5m telescopes, the last number jumps to 60-80 planets that might harbor life. Recent technical advances make space interferometry more feasible, including successful formation flying missions (e.g,. GRACE-FO, LISA Pathfinder), lower launch costs, and the maturation of cubeSats and commercial smallSats \citep[see detailed overview by][]{monnier2019_wp}. The first space interferometer SunRISE \citep[6-cubeSats in radio][]{sunrise2019} will be launched in $\sim$2024 and multiple optical designs using smallSats have been proposed \citep[e.g.,][]{matsuo2022}.  While the near-term political and technical outlook is still uncertain, the scientific potential of space nulling interferometry to measure the mid-IR spectra of Earth-like planets is still highly compelling and recognized as one of the most important long-term goals for Astronomy.

\begin{issues}[FUTURE DIRECTIONS]
\begin{enumerate}
\item Current interferometers fall many orders of magnitude behind fundamental limits.
\item Worst offenders are adaptive optics performance, optical throughput, metrology laser background, detector noise, and vibrations.
\item Ongoing GRAVITY+ upgrade with laser guide star AO and wide-field dual-beam capability will boost sky coverage by order of magnitudes.
\item Current facilities continue to expand instrumentation to wider wavelength ranges and offer a testbed for new technologies, such as nulling.
\item VLTI and CHARA will remain unique in the era of upcoming 30-40\-m extremely large telescopes (ELTs), not for sensitivity but for angular resolution.
\item New ground- and space-based interferometers are under construction (MROI) or proposed (PFI, LIFE).
\item Technologies to pursue include affordable 8-m telescopes, interferometric laser guide stars, formation flying space interferometry, and photon-counting IR detectors.
\end{enumerate}
\end{issues}

\section*{DISCLOSURE STATEMENT}
The authors are not aware of any affiliations, memberships, funding, or financial holdings that
might be perceived as affecting the objectivity of this review. 

\section*{ACKNOWLEDGMENTS}
We thank all our colleagues for inputs, discussions, comments and corrections, in particular Mark Colavita and Julien Woillez on the technical advances, and Stefan Gillessen, Sylvestre Lacour, Karine Perraut, Guy Perrin, and Taro Shimizu on the astrophysical breakthroughs. JM thanks Fabien Baron for his contributions on the interferometric imaging. FE and OP are grateful to Reinhard Genzel, Tim de Zeeuw and Dieter Lutz for many valuable inputs throughout the preparation of this manuscript. We thank our editor Jonathan Bland-Hawthorn for his feedback and guidance in preparing the review.

\bibliographystyle{ar-style2}


\end{document}